\documentclass[a4paper,11pt]{article}
\usepackage{epsfig}
\title{\bf Diffractive scattering on the deuteron projectile in the NLO: triple
interaction of reggeized gluons}
\author{M.A.Braun, S.S. Pozdnyakov, M.Yu.Salykin, M.I.Vyazovsky\\
{\it S.Petersburg State University, Russia}}
\begin{document}

\maketitle
\input epsf

\def\beq{\begin{equation}}
\def\eeq{\end{equation}}
\def\ept{\epsilon_\perp}


\def\disc{{\rm Disc}}
\def\lra{\leftrightarrow}
\def\ep{\epsilon}
\def\eps{\varepsilon}
\def\aa{a^{(1)}}
\def\ab{a^{(2)}}
\def\qa{q_{1+}}
\def\qb{q_{2+}}
\def\ra{r_{1-}}
\def\rb{r_{2-}}
\def\bat{\bar{t}}
\def\ka{\kappa_+}
\def\bk{{\bf k}}
\def\bt{{\bf t}}
\def\bp{{\bf p}}
\def\bqa{{\bf q}_1}
\def\bqb{{\bf q}_2}
\def\bra{{\bf r}_1}
\def\brb{{\bf r}_2}
\def\tka{k_{1\perp}^2}
\def\tkb{k_{2\perp}^2}
\def\im{{\rm Im}\,}
\def\re{{\rm Re}\,}
\def\ci{{\rm ci}\,}
\def\tp{p_\perp^2}
\def\tpom{{\tilde P}}
\def\th{{\tilde h}}

\def\kta{k_{1\perp}^2}
\def\ktb{k_{2\perp}^2}

\def\beq{\begin{equation}}
\def\eeq{\end{equation}}
\def\ept{\epsilon_\perp}
\def\cca{\frac{q_1^2-\lambda_\perp^2}{(q_1-\lambda_\perp)^2}}
\def\ccb{\frac{q_2^2-\lambda_\perp^2}{(q_2-\lambda_\perp)^2}}


{\bf Abstract}

\noindent
High-mass diffractive production of protons on the deuteron target is studied in the
next-to-leading order (NLO) of the perturbative QCD in
the BFKL approach.
The non-trivial part of the NLO contributions coming from the
triple interactions of the exchanged reggeons is considered.
Analytic formulas are presented and shown to be infrared free and so ready for practical calculation.

\section{Introduction}
In the perturbative QCD collisions on heavy nuclear targets have long been the object
of extensive study. In the BFKL approach the structure function of DIS on a heavy nuclear target
is given by a sum of fan diagrams in which BFKL pomerons propagate and split by the
triple pomeron vertex.
This sum satisfies the well-known Balitski-Kovchegov equation derived
earlier in different approaches ~\cite{bal,kov}. The corresponding inclusive cross-sections for gluon production
were derived in ~\cite{kov1,kov2}. Description of nucleus-nucleus collisions has met with less success.
For collision of two heavy nuclei in the framework of the Color Glass Condensate and JIMWLK approaches numerical
Monte Carlo methods were applied ~\cite{krasnitz,nara,lappi}. Analytical approaches,
however, have only given modest approximate results ~\cite{kovchegov, balitski,dusling}.
The methods developed so far refer to the collision with heavy nuclei and rely basically on the semi-classical
approximation valid in the limit of a very small coupling constant where the amplitudes behave as $1/g^2$.
To move to nuclei with the smaller number of nucleons one needs to develop an approach which explicitly
uses the composition of the nucleus as  a composite of a few nucleons which interact with the projectile
in the manner specific to the Regge kinematics. This approach is clearly provided by  the framework of the
exchange of reggeized gluons  which combine into colorless pomerons or higher BKP states
~\cite{bkp1,bkp2}(the BFKL framework).
To understand the problem one of the authors (M.A.B)
turned to the simplest case of nucleus-nucleus interaction, namely the deuteron-deuteron collisions
~\cite{braun1,braun2}. It was found that in this case the diagrams which give the leading contribution are different from the
heavy nucleus case and include non-planar diagrams subdominant in $1/N_c$ where $N_c$ is the number of colors.
In ~\cite{diflo} this technique was applied to the diffractive proton production off the deuteron and it was found that
the contribution from the diagrams involving both nucleons in the deuteron is dominant with respect to simple triple pomeron
diagrams connected to only one deuteron component. In the reggeized gluon language interaction between two pairs of
reggeized gluon dominates over such interaction with only one initial pair.

Higher orders provided by additional BFKL interactions give new terms which at large rapidities have the same order and
lead to appearance of fully developed BFKL pomerons and BKP states formed within the deuteron. However, there also appear
a true second order terms, which correspond either to second order BFKL interaction or to new elements in the
interaction involving not two pairs of reggeized gluons (reggeons) but three pairs of them. Such a NLO interaction was studied
in the odderon problem ~\cite{odderon}, where the corresponding expression were derived and, what is most important,
absence of infrared divergence was demonstrated. With that this NLO interaction became ready for the numerical analysis.

In this paper we study the NLO interactions involving three pairs of reggeons for the diffractive
production of the protons off the deuteron,  the process studied in LO in  ~\cite{diflo}. This problem is much more difficult
than a similar odderon one due to less symmetry and appearance of certain new diagrams which are absent for the
odderon just by absence of symmetry. These extra diagrams prevent using the technique of
~\cite{odderon} based on the earlier found expression for two gluon emission in ~\cite{fadin}, which substantially
simplified calculation for the odderon. Unfortunately we are bound to use a novel technique and perform rather
cumbersome calculations.
We shall discover that these extra diagrams individually contain infrared
divergency of a very unpleasant character. Our main result is to demonstrate that after summation of all numerous
contributions these infrared divergency is canceled and as for the odderon in  ~\cite{odderon} the
final expressions are ready for numerical study.

\begin{figure}
\begin{center}
\epsfig{file=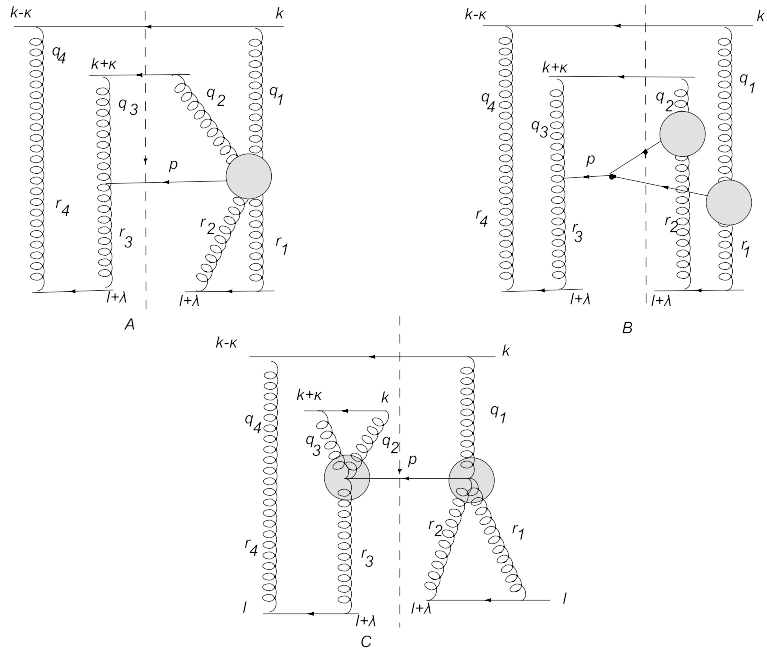, width=10 cm}
\caption{NLO diagrams with interaction of three pairs of reggeons.}
\label{fig1}
\end{center}
\end{figure}

The NLO contributions  to the diffractive production of protons off the deuteron with interaction of three pairs of reggeons
involve three types of diagrams shown in Fig \ref{fig1}. The bulk of the contribution comes from diagram
in Fig. \ref{fig1},A in which the intermediate gluon is produced by the process R+R$\to$R+R+G where R stands for
reggeon and G for gluon.  In Fig. \ref{fig1},B we show another contribution to the cross-section with two gluons in the
intermediate state. Finally Fig. \ref{fig1},C shows one more possible diagram for this process in which the intermediate gluon
arises from two processes R+R$\to$R+G. However, in fact this latter contribution is canceled between the direct and conjugated
diagrams and need not be considered.

To conclude this introduction we recall the basic formulas connecting the diagrams with the cross-section itself.
We work in the center-of-mass system of the colliding proton and one of the nucleons of the deuteron.
The inclusive cross-section of the diffractive
proton production $d(2k)+p(l)\to p(l')+X$ is given by
\beq
I(l')\equiv\frac{(2\pi)^32l'_-d\sigma}{dl'_-d^2l'_\perp}=
\frac{1}{s}{\rm Im}{\cal A},
\label{eq1}
\eeq
where the amplitude ${\cal A}$ corresponds to Fig. \ref{fig1}.
Let the final proton momentum  be $l'=l+\lambda$. The missing mass is
then $M^2=-4k_+\lambda_-$. In terms of the overall and pomeron rapidities $Y$ and $y$ we have
 $M^2=M_0^2\exp(Y-y)$ where  $M_0\sim 1$ Gev. Putting $t=|\lambda_\perp|^2$
 we rewrite the diffractive cross-section as
\beq
J(y,t)\equiv\frac{d\sigma}{dydt}=\frac{M^2}{32\pi^2s^2}\im \cal{A}.
\label{jyt}
\eeq

Separating the deuteron lines we standardly find (see~\cite{bra1})
\beq
{\cal A}=\int dz F(z)|\psi_d(r_\perp=0,z)|^2,
\label{eq2}
\eeq
where
\beq
F(z)=\frac{1}{k_+}\int\frac{d\kappa_+}{2\pi}H(\kappa_+)e^{izm\kappa_+/k_+},
\label{ffun}
\eeq
$H$ is the high-energy part of ${\cal A}$ and
$\kappa_+$ is the $+$-component of the  momentum $\kappa$
transferred to one of the nucleons in the deuteron
with $\kappa_-=\kappa_\perp=0$.
For comparison, in the same process with a heavy nucleus projectile, the
contribution from the collision with two nucleons is given by (\ref{eq1}) with
\beq
{\cal A}=\frac{1}{4}A(A-1)\int d^2bdz_1dz_2 F(z_1-z_2)\rho({\bf b},z_1)\rho({\bf b},z_2),
\label{anuc}
\eeq
where $\rho({\bf b},z_1)$ is the nuclear density normalized   to unity.

The Glauber approximation corresponds to the contribution
which follows when $F(z)$ does not depend on $z$. Then the square of the deuteron
wave function converts into the average $<1/2\pi r^2>$ and in (\ref{anuc}) we find
integration over the impact parameter ${\bf b}$ of the square of the profile function
$T({\bf b})$. In standard cases the high-energy part contains $\delta(\kappa_+)$:
\beq
H(\kappa_+)=2\pi \delta(\kappa_+)D,\ \ {\rm so\ that}\ \
F=\frac{1}{k_+}D.
\eeq
Then for the deuteron
\beq
{\cal A}=\frac{D}{k_+}<1/2\pi r^2>_d
\label{ad}
\eeq
and for a large nucleus
\beq
{\cal A}=\frac{1}{4}A(A-1)\frac{D}{k_+}\int d^2bT^2(\bf b).
\label{aa}
\eeq

The paper is organized as follows. In the next section we discuss
the main part of the contribution to the high-energy amplitude $H$
corresponding to the transition R+R$\to$R+R+G for the production
of the intermediate gluon realized by vertex $\Gamma_{RR\to RRG}$.
Next we discuss the two-gluon intermediate state corresponding
to Fig. \ref{fig1},B. In the last section we make some conclusions.
Some long and cumbersome calculations are transferred to the three
Appendices.

\section{Contribution from the RR$\to$RRG vertex}
The  diagram which describes the  NLO corrections due to RR$\to$RRG vertex $\Gamma_{RR\to RRG}$
is shown in Fig. \ref{fig1},A in the introduction. It should be supplemented by a similar diagram with interchanged
nucleons in the deuterons and conjugated contributions. The interchange of the nucleons does not change the amplitude.
due to the symmetry of the vertex respective to permutations of both the two incoming reggeons and the outgoing reggeons.
So it is sufficient to study the diagram in Fig. \ref{fig1},A and double its contribution.
Vertex  $\Gamma_{RR\to RRG}$ itself does not depend on impact factors and does not feel evolution.  So finding its contribution can be
simplified by suppressing evolution and taking some simple impact factors for the pomerons.
We take simple quarks for the four scattering centers
assuming that their interaction is due to colorless exchange.

The total number of transferred momenta is 7: $q_1$, $q_2$, $q_3$, $r_1$, $r_2$,
$r_3$, and $q_4=r_4$.
The momentum of the real gluon $p$ is  $p=q_1+q_2-r_1-r_2$.
We choose as independent momenta $q_1$, $r_1$ and $p$. Then we have
\[q_2=p+\lambda-q_1,\ \ q_3=q_1-p-\kappa-\lambda,\ \ q_4=r_4=\kappa-q_1,\ \
r_2=\lambda-r_1,\ \ r_3=q_1-\kappa-\lambda.\]
So we have 6 longitudinal integrations.
There are 5  conditions arising from mass-shell conditions for real intermediate particles
and sums of direct and crossed diagrams for the rest, which give:
\[
(2\pi)^5\delta(q_{1-})\delta(q_{2-})\delta(r_{1+})\delta(r_{4+})\delta(p^2)
\]
multiplied by $4s^2$.
Integration over $q_{1+}$, $q_{1-}$ and $r_{1+}$  are done withe the help of $\delta$-functions, which
puts $q_{1+}=\kappa_+$, $q_{1-}=r_{1+}=0$. Of the three  integrations
over $p_{\pm}$ and $r_{1-}$
the $\delta$ functions
\[
(2\pi)^2\delta(p^2)\delta(p_-+\lambda_-)
\]
allow to integrate over $p_{\pm}$ and we are finally left with only one
longitudinal integration over $\ra$
with $p_-=-\lambda_-$.

Apart from the four pomerons the diagram of Fig. \ref{fig1},A
involves the Lipatov vertex on the left
$ f^{b_3a_3c}L(-p,r_3)$
where $a_3$, $b_3$ and $c$ are the color indices of the two
reggeons 3 (incoming and out going) and of the real gluon. It does not depend
on longitudinal variables and enters only the transversal integral.
On the right we meet the RR$\to$RRP vertex of the structure
\[ \Gamma^{a_2a_1c}(q_2,q_1|r_2,r_1)=f^{a_2a_1c}\Big(\Gamma(q_2,q_1|r_2,r_1)-\Gamma(q_1,q_2|r_2,r_1)\Big).\]
It does depend on longitudinal variables and is symmetric  in $r_1,r_2$ and antisymmetric
in $q_1,q_2$.

Doing summation over colors
we obtain the imaginary part $H_1$ coming from Fig. \ref{fig1},A as
\[{\rm Im}\, H_1=
g^4\frac{2s^2}{ p_-}\int \frac{dr_{1-}}{2\pi}\int d\tau_\perp
 P_{Y-y}(q_{1\perp},q_{4\perp})P_{Y-y}(q_{2\perp},q_{3\perp})L(r_3-q_3,r_3)\]\beq
 {\rm Im}\,\Big(
\Gamma_(q_2,q_1|r_2,r_1)-\Gamma(q_1,q_2|r_2,r_1)\Big)
P_{y}(r_{1\perp},r_{2\perp})P_{y}(r_{3\perp},r_{4\perp}),
\label{imh1}
\eeq
where $\tau_\perp$ is the transverse phase volume, $y$ is defined as before via $M^2$.
We include in each pomeron factor $N_cg^2$. Then in the final factor  just $g^4$ appears.
Note that $2s^2/p_-=8s^2k_+/M^2$.

In (\ref{imh1}) we have used that the momentum part of $\Gamma$ is antisymmetric under $\qa\lra\qb$, since the color factor
is antisymmetric under this exchange.
In  the vertex $\Gamma$ we have to take longitudinal
variables in accordance with our results
\[q_{1+}=\kappa_+,\ \  q_{1-}=0,\ \ q_{2+}=p_+-\kappa_+,\ \ q_{2-}=0,\ \ r_{1+}=0,\ \  r_{2+}=0,\]\[
r_{2-}=\lambda_{-}-r_{1-},\ \  p_-=-\lambda_-,\ \ p^2=0\]
and the transverse momenta inside the pomerons are constrained by
\beq
 q_{1\perp}+q_{4\perp}=q_{2\perp}+q_{3\perp}=0,\
\ r_{1\perp}+r_{2\perp}=-r_{3\perp}-r_{4\perp}=\lambda_{\perp}=l'_\perp.
\label{tkinrel}
\eeq
The interchange of the two projectiles in the deuteron plus complex conjugate contribution multiply
(\ref{imh1}) by factor 4.

Note that in contrast to more or less trivial cases of the scattering off a composite the
obtained expression does not contain $\delta(\kappa_+)$, which could lift the integration in
(\ref{ffun}) and bring the resulting cross-section into the Glauber form. In our case
$\kappa_+$ appears in the cross-section via longitudinal momentum $\qa=\kappa_+$. Therefore
in (\ref{imh1}) after doing the integration over $\ra$ one has to perform integration over $\qa$
with the weight dictated by (\ref{ffun}). As we shall see this integration goes over a finite interval
of rapidities of the order $\delta$ due to  the condition that all intermediate gluons should lie at finite rapidity distances from
the real intermediate gluon.
After integration over $\qa$ one finds first terms  proportional to $\delta$, which should be dropped, since hopefully
they are to be canceled by other terms of the same order,
namely  terms with pairwise interaction between reggeons, with extra BFKL interactions or with this interaction
in the second order. The rest terms are well convergent and  do not depend on $\delta$
nor on the exponential in (\ref{ffun}), since the exponent is small. These terms lead to $F(z)$ independent of $z$
and thus to the same Glauber expression (\ref{ad}) for the amplitude  with
\beq
F=\frac{1}{k_+}\int\frac{d\qa}{2\pi}\im H.
\label{dnew}
\eeq

As mentioned, the left Lipatov vertex in the diagram in Fig. \ref{fig1},A is on-shell
and does not depend on $r_{1-}$ nor on $\qa$
So we have to longitudinally integrate only the vertex RR$\to$RRP.
Therefore
combining the coefficients in (\ref{imh1}) and (\ref{jyt})  our result final result for the cross-section from the vertex $\Gamma$
has the form
\[
\frac{d\sigma_{\Gamma}}{dydt}=<1/2\pi r^2>_d\frac{g^4N_c^4}{\pi^2}
\int d\tau_\perp {\cal M}\times\]\beq\times
 P_{Y-y}(q_{1\perp},q_{4\perp})P_{Y-y}(q_{2\perp},q_{3\perp})
P_{y}(r_{1\perp},r_{2\perp})P_{y}(r_{3\perp},r_{4\perp}),
\label{crsecg}
\eeq
where
\beq
{\cal M}=L(r_3-q_3,r_3){\rm Im}\,\int\frac{d\kappa_+d\ra}{4\pi^2}
\Big(\Gamma(q_2,q_1|r_2,r_1)-\Gamma(q_1,q_2|r_2,r_1)\Big) .
\label{mm}
\eeq

\begin{figure}
\begin{center}
\epsfig{file=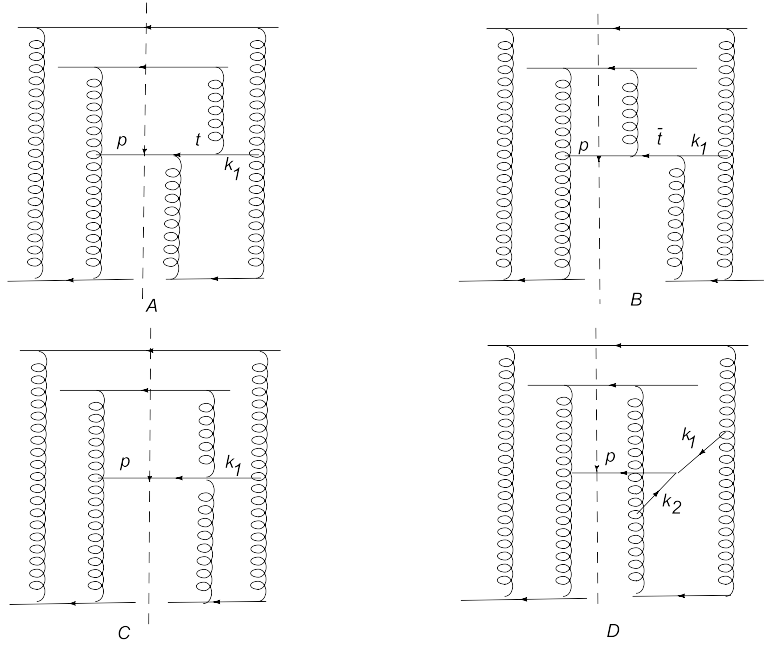, width=10 cm}
\caption{Diagrams containing vertex parts $\Gamma_i$, $i=1,..5$.
$\Gamma_{1,2,5}$ correspond to A,B and D. $\Gamma_{3,4}$ correspond to C.}
\label{fig2}
\end{center}
\end{figure}

The total vertex $\Gamma$ is a sum of 5 pieces
$\Gamma=\sum_{i=1}^5 \Gamma_i$.
The diagrams with vertices $\Gamma_i$ are shown in Fig. \ref{fig2}.
Actual longitudinal integrations in $\Gamma_i$, $i=1,...5$ are long and tedious. They are described in  Appendices 1,2 and 3.
To avoid proliferation of notations we denote the resulting integrated $\Gamma$ with the same letter $\Gamma$.
Collecting our results derived  in the appendices,
we find  for different pieces the following expressions.

\[\Gamma_1=\frac{i}{8\pi}\hat{T}_0\frac{1}{2(\bp+\brb)^2}\ln\frac{(\bqa-\bra)^2\bp^2}{(\bp+\brb)^4}+\Big(Sym\Big),
\]
where $\hat{T}_0$ is given by (\ref{ht10}).

\[\Gamma_2=-\frac{i}{4\pi}\hat{T}_1\frac{1}{2\bp^2}\ln\frac{(\bqa-\bra)^2\bp^2}{(\bp-\bqb)^4}+
\Big(Sym\Big),\]
where $\hat{T}_1$ is given by (\ref{ht21}).

\[\Gamma_3=\frac{i}{4\pi}V_0\frac{1}{2(\bqa-\bra)^2}\ln\frac{(\bqa-\bra)^2}{\bp^2}+\Big(Sym\Big).\]
where $V_0$ is given by (\ref{v03}).

\[\Gamma_4=-\frac{i}{8\pi}V_1\frac{1}{2\bp^2}\ln\frac{(\bqa-\bra)^2}{\bp^2}+\Big(Sym\Big),
\]
where $V_1$ is given by (\ref{v14}).
Here $+(Sym)$ means addition of
\[-\Big(q_1\lra q_2\Big)+
\Big(r_1\lra r_2\Big)-\Big(q_1\lra q_2, r_1\lra r_2\Big)\].

The most complicated part comes from $\Gamma_5=\Gamma^A+\Gamma^B$
We find
\[
\Gamma^A=\frac{i}{8\pi}\Big\{I^A_0\Big[T_2\Big(2\frac{(\bp,\bqb-\brb)^2}{\bp^4}-\frac{(\bqb-\brb)^2}{\bp^2}\Big)+
T_1\frac{(\bp,\bqb-\brb)}{\bp^2}+T_0
+T_{-1}\frac{(\bp,\bqb-\brb)}{(\bqb-\brb)^2}\]\[+\frac{U_0}{p_+}\frac{(\bp,\bqa-\bra)}{(\bqa-\bra)^2}
-\frac{V_0}{p_+}\frac{(\bqa-\bra,\bqb-\brb)}{(\bqa-\bra)^2(\bqb-\brb)^2}\Big]\]
\[
-\frac{1}{2(\bqa-\bra)^2}\ln\frac{(\bqb-\brb)^2}{\bp^2}\Big(\frac{U_0}{p_+}-\frac{1}{(\bqb-\brb)^2}\,\frac{V_0}{p_+}\Big)\Big\} -\Big(q_1\lra q_2\Big),\]
where
\[
I_0^A=\frac{\pi-\phi_2}{\sqrt{\bp^2(\bqb-\brb)^2-(\bp,\bqb-\brb)^2}},
\]
$\phi_2$ is the angle between $\bp$ and $\bqb-\brb$, $0\leq\phi\leq\pi$ and coefficients $T_n$, $U_0$ and $V_0$ are
given by (\ref{tuva}).

\[
\Gamma^B=-\frac{i}{8\pi}\Big\{I^B_0\Big[\tilde{T}_2\Big(2\frac{(\bp,\bqa-\bra)^2}{\bp^4}-\frac{(\bqa-\bra)^2}{\bp^2}\Big)+
\tilde{T}_1\frac{(\bp,\bqa-\bra)}{\bp^2}+\tilde{T}_0+\tilde{T}_{-1}\frac{(\bp,\bqa-\bra)}{(\bqa-\bra)^2}\]\[+\frac{\tilde{U}_0}{p_+}\frac{(\bp,\bqb-\brb)}{(\bqb-\brb)^2}
-\frac{\tilde{V}_0}{p_+}\frac{(\bqa-\bra,\bqb-\brb)}{(\bqa-\bra)^2(\bqb-\brb)^2}\Big]\]
\[
+\frac{1}{2(\bqb-\brb)^2}\ln\frac{(\bqa-\bra)^2}{\bp^2}\Big(\frac{\tilde{U}_0}{p_+}-\frac{1}{(\bqa-\bra)^2}\,\frac{\tilde{V}_0}{p_+}\Big)\Big\} -\Big(q_1\lra q_2\Big),\]

where
\[
I_0^B=\frac{\phi_1}{\sqrt{\bp^2(\bqa-\bra)^2-(\bp,\bqa-\bra)^2}},\]
$\phi_1$ is the angle between $\bp$ and $\bqa-\bra$, $0\leq\phi\leq\pi$ and coefficients $\tilde{T}_n$, $\tilde{U}_0$ and $\tilde{V}_0$ are
given by (\ref{tuvb}).

Separate terms in $\Gamma_3$, $\Gamma_4$ and $\Gamma_5$ contain non-integrable divergence at $\bqa\to\bra,\brb$ and $\bqb\to\bra,\brb$.
However, in Sec. 7.6 it is demonstrated that these singularities cancel in the sum of all $\Gamma_i$, $i=3,4,5$.

In (\ref{mm})  the integrated $\Gamma$ is to be multiplied by the Lipatov vertex
\[
L(r_3-q_3,r_3)=-2\Big((q_2e)_\perp-(pe)_\perp\frac{q_2^2}{p_\perp^2}\Big).\]
The integrated $\Gamma$ contains products $(ae)_\perp$ with different $a$. Summation over polarization
transforms
\[L(r_3-q_3,r_3)(ae)\to 2\Big((q_2a)_\perp-(pa)_\perp\frac{q_2^2}{p_\perp^2}\Big).\]
So in the end one obtains ${\cal M}$ as a well defined function of transverse momenta ready for practical evaluations.

\section{Two intermediate gluons}
To begin we find that of the two additional contributions shown in Fig. \ref{fig1}, B  and C only the first with two intermediate
gluons gives non-zero contribution. Indeed the contribution to the unitarian from the diagram in
Fig. \ref{fig1}, C  cancels between the amplitude and its conjugated
term.   In this diagram on the right-hand side (rhs) from the cut  we find purely imaginary contribution due to the single
incoming reggeon $q_1$. On the left-hand side (lhs)  we have  a similar purely imaginary quantity plus an extra gluon $q_4=r_4$,
which gives $+i$. So the total contribution  to the unitarian is imaginary and will be canceled by the
conjugated one.

So we are left only with the contribution from the two-gluon intermediate states, Fig. \ref{fig1}, B.
This diagram is quite similar to the diagram in Fig. \ref{fig2},D but with a different cut. This makes its calculation
somewhat different.

\subsection{Longitudinal integration}
The cut separates the diagram into two parts: lhs and rhs.
For the lhs we find
\beq
{\rm lhs}=\frac{1}{p^2+i0}\Big[2(k_1e_2)(Le_1)-2(k_2e_1(Le_2)+(e_1e_2)(L(k_2-k_1))\Big],
\label{lhs}
\eeq
where we denoted $L=L(-p,r_3)$ and $e_1=e(k_1)$ and $e_2=e(k_2)$ are the 4-dimensional
polarization vectors with $e_+=0$.
For the rhs we have
\beq
{\rm rhs}=16\pi^2\delta(k_1^2)\delta(k_2^2)
\Big(q_1e_1)_\perp-(k_1e_1)\frac{q_1^2}{\kta}\Big)
\Big(q_2e_2)_\perp-(k_2e_2)\frac{q_2^2}{\ktb}\Big).
\eeq

Integrations over $q_{1+}$ and $q_{2_+}$ are done using the two $\delta$ functions.
As a result we get factor $1/(4r_{1-}r_{2-})$ and $ q_{1,2+}$ become expressed as
\beq q_{1+}=\frac{\kta}{2\ra},\ \  q_{2+}=\frac{\ktb}{2\rb}.
\label{qviar}
\eeq

The final longitudinal integration is over $r_{1-}$ with $r_{2-}=\lambda_--r_{1-}$
In the denominator appears
\[
D=r_{1-}r_{2-}p^2=r_{1-}r_{2-}\Big[-2\lambda_-\Big(\frac{\kta}{2r_{1-}}+
\frac{\ktb}{2\rb}\Big)+p_\perp^2\Big]=
\]\beq
-\lambda_-\Big(k_{1\perp}^2r_{2-}+k_{2\perp}^2r_{1-}\Big)+r_{1-}r_{2-}p_\perp^2.
\eeq
Putting $r_{1-}=x\lambda_-$ and so $r_{2-}=(1-x)\lambda_-$ we rewrite
\beq
D=-\lambda_-^2\Big((1-x)k_{1\perp}^2+xk_{2\perp}^2-x(1-x)p_\perp^2\Big)
=-\lambda_-^2(k_{1\perp}-xp_\perp)^2.
\label{den}
\eeq
So we see that the contribution from diagram Fig. \ref{fig1},B contains the
same denominator as the contribution from $\Gamma_5$ (see Appendix). However, as we discuss in Sec. 7.5  the
collinear singularity from $D=0$ is spurious, since the numerator vanishes.

Rhs does not depend on $r_{1-}$. Lhs contains factors depending on $\ra$
Making explicit the $x$-dependence  we have in lhs
\[ (k_1e_2)=(k_1e_2)_\perp-(k_2e_2)_\perp\frac{1-x}{x}\frac{\kta}{\ktb},\]
\[ (k_2e_1)=(k_2e_1)_\perp-(k_1e_1)_\perp\frac{x}{1-x}\frac{\ktb}{\kta},\]
\[(Le_1)=(b_1,e_1)_\perp-(k_1e_1)_\perp\Big(\frac{x}{1-x}\frac{\ktb}{\kta}-2x\frac{q_3^2}{\kta}\Big),\]
\[(Le_2)=(b_2,e_2)_\perp-(k_2e_2)_\perp\Big(\frac{1-x}{x}\frac{\kta}{\ktb}-2x\frac{q_3^2}{\ktb}\Big).\]
\beq
(L,k_2-k_1)=-\kta r_3^2\frac{1-x}{(1-x)\kta+x\ktb}-\frac{1}{2}\kta\frac{1-x}{x}
+\frac{1}{2}\ktb\frac{x}{1-x}+xq_2^2+(k_1,q_3+r_3)_\perp
\label{lk}
\eeq
Naturally this expression changes sign if $k_{1\perp}\lra k_{2\perp}$ and $x\lra 1-x$. The form
(\ref{lk})  is convenient for the study of the limit $k_{1\perp}\to 0$.
Here $b_1=q_3+r_3-k_1$, $b_2=q_3+r_3-k_2$.
So lhs contains singular factors $1/x$ and $1/(1-x)$ and grows linearly with $x$ at large $x$. The singularities at $x=0$ and $x=1$
are to be integrated in the principal value sense. At large $x$ the integrand reduces to $1/x$ and the integral over the whole
axis converges.

The longitudinal integral over $r_{1-}$ takes the form
\beq
J=-\int\frac{dx}{8\pi\lambda_-}\frac{X_1-X_2+X_3}{(k_{1\perp}-xp_\perp)^2}.
\eeq
We define transverse vectors
\[l_1=q_1-k_1\frac{q_1^2}{\kta},\ \ l_2=q_2-k_2\frac{q_2^2}{\ktb},\ \ l_{1\pm}=l_{2\pm}=0.\]
Then after summation over polarizations we get
\beq
X_1=8\Big(l_2, k_1-k_2\frac{1-x}{x}\frac{\kta}{\ktb}\Big)_\perp
\Big[l_1,b_1-k_1\frac{x}{1-x}\frac{\ktb}{\kta}\Big(1-2(1-x)\frac{q_3^2}{\ktb}\Big)\Big]_\perp,
\label{x1}
\eeq
\beq
X_2=8\Big(l_1, k_2-k_1\frac{x}{1-x}\frac{\ktb}{\kta}\Big)_\perp
\Big[l_2,b_2-k_2\frac{1-x}{x}\frac{\kta}{\ktb}\Big(1-2x\frac{q_3^2}{\kta}\Big)\Big]_\perp,
\label{x2}
\eeq
\beq
X_3=(L,k_2-k_1)(l_1l_2),
\label{x3}
\eeq
where $(L,k_2-k_1)$ is given by (\ref{lk}).
Doing the products in (\ref{x1}) and (\ref{x2}) we rewrite them in terms of transverse products
\[
\frac{1}{8}X_1=(k_1l_2)(b_1l_1)-(k_2l_2)(b_1l_1)\frac{1-x}{x}\frac{\kta}{\ktb}\]\beq
-(k_1l_2)(k_1l_1)\frac{x}{1-x}\frac{\ktb}{\kta}\Big(1-2(1-x)\frac{q_3^2}{\ktb}\Big)
+(k_2l_2)(k_1l_1)\Big(1-2(1-x)\frac{q_3^2}{\ktb}\Big),
\label{x11}
\eeq
\[
\frac{1}{8}X_2=(k_2l_1)(b_2l_2)-(k_1l_1)(b_2l_2)\frac{x}{1-x}\frac{\ktb}{\kta}\]\beq
-(k_2l_1)(k_2l_2)\frac{1-x}{x}\frac{\kta}{\ktb}\Big(1-2x\frac{q_3^2}{\kta}\Big)
+(k_2l_2)(k_1l_1)\Big(1-2x\frac{q_3^2}{\kta}\Big).
\label{x12}
\eeq

The integrals over $x$ are all standard. With $(k_{1\perp}-xp_\perp)^2\equiv d$
we have
\[
I=\int\frac{dx}{d}=\frac{\pi}{\sqrt{k_1^2p^2-(k_1p)^2}}=\frac{\pi}{\sqrt{k_2^2p^2-(k_2p)^2}},
\]
\[
\int\frac{dx}{xd}=\frac{(k_1p)}{\kta}I,\ \ \int\frac{dx}{(1-x)d}=\frac{(k_2p)}{\ktb}I,\]
\[
\int\frac{xdx}{d}=\frac{k_1p}{p^2}I,\ \ \int\frac{(1-x)dx}{d}=\frac{k_2p}{p^2}I,\]
\[
\int\frac{xdx}{(1-x)d}=\frac{(k_1k_2)}{\ktb}I,\ \ \int\frac{(1-x)dx}{xd}=\frac{(k_1k_2)}{\kta}I
\]
and finally
\[
I_1=\int\frac{dx}{d}\frac{(1-x)\kta-x\ktb}{(1-x)\kta+x\ktb}= \frac{k_1^2-k_2^2}{p^2}I\]
(antisymmetric under $k_1\lra k_2$).
In these and the following formulas all vectors are 2-dimensional Euclidean.

Using them we finally find for the integrated quantities
\[
Z_1=\int\frac{dx}{d}X_1=8I\Big\{(k_1l_2)(b_1l_1)-(k_2l_2)(b_1l_1)\frac{(k_1k_2)}{\ktb}\]\[
-(k_1l_2)(k_1l_1)\frac{(k_1k_2)}{\kta}+2(k_1l_2)(k_1l_1)\frac{(k_1p)}{\kta}\,\frac{q_3^2}{p^2}\]\beq
+(k_2l_2)(k_1l_1)-2(k_2l_2)(k_1l_1)\frac{(k_2p)}{\ktb}\,\frac{q_3^2}{p^2}\Big\},
\label{z1}
\eeq
\beq
Z_2=\int\frac{dx}{d}X_2=Z_1(k_1\lra k_2)
\label{z2}
\eeq
and
\beq
Z_3=\int\frac{dx}{d}X_3=I(l_1l_2)\Big\{(k_2^2-k_1^2)\Big(1-\frac{r_3^2}{p^2}\Big)+k_2^2-k_1^2+(q_3+r_3,k_2-k_1)\Big\}.
\label{z3}
\eeq

\subsection{The cross-section}
Apart from factor $Z_1-Z_2+Z_3$ the contribution to the high-energy part
will include color, longitudinal and transverse factors, which can be readily read from the diagram.

The final longitudinal factor comes from  $-1/\lambda_-$ in $D$ and factors $2k_+$ from each projectile quark
and $2l_-$ from each target one, which gives the total $-16k_+^2l_-^2/\lambda_-$.
The color factor $(1/2)N_c^4$ is the same as in Fig. \ref{fig2},D.
\[ f^{a_2a_1c}f^{ce_2e_1}f^{e_1b_1a_1}f^{e_1b_1a_2}=\frac{1}{2}N_c^4.\]

The transverse factor $T$ is obtained after transverse integration of the sum $Z_1-Z_2+Z_3$ with the
pomerons coupled to the projectiles and targets
\beq
T(y,t)=\int d\tau_\perp(Z_1-Z_2+Z_3)P_{Y-y}(q_1)P_{Y-y}(q_2)
P_y(q_1,q_1-\lambda)P_y(q_1,\lambda-q_1),
\label{t}
\eeq
where $t=-\lambda_\perp^2$.
As before we include in each pomeron factor $N_cg^2$. Then the final extra factor will be just $g^4$.

Note that  $Z_i$ $i=1,2,3$ contain singular terms. They first come from function $I(k_1,k_2)$, which is singular
when $k_{1\perp}$ is parallel to $p_\perp$ (collinear singularity) and also when one of the 2-dimensional vectors
$k_1,k_2$ or $p$ goes to zero. The first singularity goes, since the coefficient vanishes as the 4-vectors become
lying in the same direction. The second singularity is integrable by itself but it may be accompanied by explicit singularities
in $Z$'s which  have the structure $(kp)_\perp/k^2$ where $k$ is any of the two dimensional vectors
$k_{1\perp},\ \ k_{2\perp}$ or  $p_\perp$.
This combined singularity is canceled after averaging over angles
and all the rest singularities turn out to be integrable.

To see this we present expressions for $Z_i$ in the limit $k_{1\perp}\to 0$.
From (\ref{z1}) we find directly
\beq
Z_1(k_{1\perp}\to 0)=-8q_1^2I\Big[(pl_2)\Big(1-\frac{2q_2^2+(pk_1)_\perp}{\tp}\Big)
+(kl_2)_\perp\Big(\frac{(pk_1)_\perp}{\kta}+1-2\frac{(pk_1)_\perp q_2^2}{p_\perp^4}\Big)\Big].
\label{z10}
\eeq
For $Z_2$ we find after making the change $k_{1\perp}\lra k_{2\perp}$
\beq
Z_2(k_{1\perp}\to 0)=-8Iq_1^2(pl_2)_\perp\frac{(pk_1)_\perp}{\kta}\Big(1-\frac{(pk_1)_\perp}{\tp}\Big).
\label{z20}
\eeq
Finally
\beq
Z_3(k_{1\perp}\to 0)=\frac{(pk_1)_\perp}{\kta}q_1^2\Big((pk_1)_\perp\frac{r_3^2-q_2^2}{\tp}-(r_3-q_2,k_1)_\perp\Big).
\label {z30}
\eeq
Inspecting these expressions we see that in the limit $k_{1\perp}\to 0$, apart from integral $I$,
$Z_1$ and $Z_3$ remain finite and $Z_2$ has a singularity proportional to
$(pk_1)_\perp/\kta$. This latter singularity is, as mentioned, liquidated after
integration over the angle between $k_{1\perp}$ and $p_\perp$. So in the end
the only remaining singularity is in $I$ and it is integrable.

Dividing by 2 to have the imaginary part we finally find for the diagram
\beq
\im H=-g^4\frac{s^2}{4\lambda_-}T(y,t).
\eeq
The contribution to the cross-section will be given by
\beq
\frac{d\sigma}{dydt}
=\frac{1}{8}\alpha_s^2T(y,t)<1/2\pi r^2>_d.
\label {j}
\eeq

The total contribution will be given by twice the sum of the contributions given by the
diagram in Fig. \ref{fig2}B and the diagram with interchange of gluons $1\lra 2$.
We take into account that the color factor and terms $Z_i$, $i=1,2,3$ change sign under this
interchange. So the net result will be symmetrization of the pomeron part.
Thus we find the cross-section as
\beq
\frac{d\sigma_{2gluon}}{dydt}
=\frac{1}{4}\alpha_s^2T^{tot}(y,t)<1/2\pi r^2>_d,
\label{crsec2}
\eeq
where
\[
T^{tot}(y,t)=\int\frac{d^2q_1d^2q_2d^2r_1}{(2\pi)^6}(Z_1-Z_2+Z_3)\]\beq\times
\Big(P_{Y-y}(q_1)P_{Y-y}(q_2)P_y(q_1,q_1-\lambda)P_y(q_1,\lambda-q_1)+(q_1\lra q_2)\Big).
\label{ttot}
\eeq

\section{Conclusions}
We considered the high-mass diffraction on the deuteron in the perturbative QCD reggeon
(BFKL-Barters) framework. It has already been shown in ~\cite{diflo} that interaction with both
components in the deuteron leads to the cross-section which may dominate over the naive triple-pomeron
contribution. In this paper we study the NLO contributions due to the novel structure appearing in the next order
and describing the triple interactions between the exchanged reggeons. The corresponding cross-sections
are presented in Eqs. (\ref{crsecg}) and (\ref{crsec2}). The important result found in relation to these cross-sections is
the demonstration that they are free from infrared divergencies and so fit for the practical evaluation.

As to these practical calculations we have to stress that the found NLO corrections are not the only one. Another
contribution comes from the 2nd order BFKL interaction in the diagrams considered in ~\cite{diflo}.
Unfortunately one cannot use for them the results found in the study of similar correction to the BFKL equation
neither in the vacuum nor octet channels, since the color structure is different in our case. Thus
the study of this particular correction requires a new derivation, which, as well-known, is quite long and complicated.
Because of this we postpone it to the future separate publication.

Finally we have to note that before attempting to perform practical calculations in the NLO one should
try to find corrections to the LO due to appearance of the BKP states in the course of evolution.
Their behavior at large energies is known and is subdominant with respect to BFKL pomeron. So one
may hope that their influence is also subdominant.  However, to make some concrete estimates one should be
able to present their wave functions in some possibly approximate form admitting practical use.
This point is also to be studied later.

\section{Appendix 1. Integration over $\ra$}
\subsection{General rules}
Variable $r_{1-}$ enters one or two Feynman denominators and also
may appear in the numerators.
One can find (numerically) that at fixed $p$ the on-mass shell vertex
multiplied by the polarization vector $\ep$ with $\ep_+=0$
goes as $1/r_{1-}^2$ as $r_{1-}\to \infty$ ~\cite{bpsv}. This allows to perform the
integration over $r_{1-}$ by closing the contour in the complex plane
and taking residues at poles.

In reality the integrand is a sum of terms $\Gamma_i$ with $i=1,...5$
which individually do not go to zero at $\ra\to\infty$ and contain
Feynman poles as well as poles at $\ra=0$ or $\rb=0$.
Since we know that  sum of these terms goes to zero at large $\ra$ fast enough,
we can forget about the behavior of individual contributions at
$\ra\to \infty$ and
and just take the residues  in, say, the lower half plane.
However, it is important that in all diagrams the residues are to be taken in the same
(lower) half plane.
Using these rules we can do integrations in  terms $\Gamma_i$
with $i=1,...5$ separately. Explicit expressions for the vertex RR$\to$RRP can be taken from our paper~\cite{bpsv}.

\subsection{Fig. \ref{fig2},A}
On mass shell multiplied by the polarization vector $\ep$ the corresponding amplitude $\Gamma_{1}$ is
given by
\beq
\Gamma_{1}=-C_1\frac{1}{t^2k_1^2}X_1,\ \
X_1=-b\bar{B}-c\bar{C}+e\bar{E}.
\label{gam1}
\eeq
Here the denominators are
$t^2k_1^2=(-2p_+r_{1-}+t_\perp^2+i0)(-2\qa r_{1-}+k_{1\perp}^2+i0)$.
The color coefficient is
 $C_1=-(1/2)N_cf^{a_2a_1c}$
The coefficients $b,c$ and $e$ are
\[b=2p_+\Big((q_1\ep)_\perp-(p\ep)_\perp\frac{\qa}{p_+}\Big)-
2\qa(r_2\ep)_\perp,\]
\[c=2p_+\Big((q_2\ep)_\perp-(p\ep)_\perp\frac{\qb}{p_+}\Big)-
2\qb(r_2\ep)_\perp,\]
\[e=-2(p+r_2,\ep)_\perp=-2(t\ep)_\perp.\]
They  do not depend on $\ra$ nor on $\rb$
We finally have
\[\bar{B}=-4\ra,\ \ \bar{C}=-4\ra+2\frac{r_1^2}{\qa},\]
\[\bar{E}=-2\ra(2\qa+\qb)+q_1^2+q_2^2-k_1^2+r_1^2+(a_1,t+q_2)_\perp
+2r_1^2\frac{\qb}{\qa}-\frac{r_1^2q_2^2}{\qa\ra},\]
where $a_1=q_1+r_1$.
From this  we find
\[X_1=-4(q_2\ep)_{\perp}p_+\frac{r_1^2}{\qa}-2(t\ep)_\perp\Big(r_1^2+q_1^2+q_2^2-k_1^2+(a_1,t+q_2)_\perp)\Big)\]\[
+4\ra p_+(t+2r_1,\ep)_\perp +4\ra\qa(t\ep)_\perp+2(t\ep)_\perp\frac{q_2^2r_1^2}{\qa\ra}.\]

We separate the term with a pole at $\ra=0$
presenting
\[X_1=\tilde{X}_1+\frac{1}{\ra}Y_1,\ \ Y_1=2(te)_\perp\frac{r_1^2q_2^2}{q_{1+}},\]
where $Y_1$ does not depend on $\ra$.

 After integration we get
 \[\Gamma_{1}=-C_1\Big(I_{A_1}^{(1)}X_1^{(1)}+I_{A_1}^{(2)}Y_1\Big).\]
 Here $X_1^{(1)}=X_1(\ra=k_{1\perp}^2/2\qa)$ and the integrals are
 \beq
I_{A_1}^{(1)}=\int\frac{dr_{1-}}{2\pi}\frac{1}{(t^2+i0)(k_1^2+i0)}
=-\frac{i}{2}\theta(-\qa)\frac{1}{p_+k_{1\perp}^2-\qa t_\perp^2}
\eeq
and $I_{A_1}^{(2)}$ is the contribution of the residue at $\ra=0$ in the lower half plane of the part
with $Y_1$
\[
I_{A_1}^{(2)}=-\frac{i}{2}\frac{1}{k_{1\perp}^2 t_\perp^2}.
\]

\subsection{ Fig. \ref{fig2},B}
On mass shell multiplied by the polarization vector $\ep$ the corresponding amplitude $\Gamma_{2}$ is
given by
\beq
\Gamma_2=-C_2\frac{1}{\bar{t}^2k_1^2}X_2,\ \ X_2=\bar{a} A+
\bar{b} B+\bar{c} C+\bar{e} E.
\eeq
Here the color factor is
$ C_2=N_c f^{a_2a_1c}$ and the denominators are
\[\bar{t}^2k_1^2=(2\qa p_-+\bar{t}_\perp^2)(-2\qa r_{1-}+k_{1\perp}^2+i0).\].
The coefficients $\bar{a},...\bar{e}$ are
\[\bar{a}=(p\ep)_\perp\frac{\bar{t}^2}{p_+}\ \  {\rm where}\ \
\bar{t}^2=2\qa p_-+\bar{t}_\perp^2,\]
\[\bar{b}=2p_-(r_1\ep)_\perp+2\ra\Big((q_2\ep)_\perp-
(p\ep)_\perp\frac{\qb}{p_+}\Big)+2(p\ep)_\perp
\Big(\ra-\ra\frac{q_2^2}{p_\perp^2}+\frac{(pr_1)_\perp}{p_+}\Big),\]
\[\bar{c}=2p_-(r_2\ep)_\perp+2\rb\Big((q_2\ep)_\perp-
(p\ep)_\perp\frac{\qb}{p_+}\Big)+2(p\ep)_\perp
\Big(\rb-\rb\frac{q_2^2}{p_\perp^2}+\frac{(pr_2)_\perp}{p_+}\Big),\]
\[\bar{e}=2(q_2\ep)_\perp+2(p\ep)_\perp\Big(1-\frac{\qb}{p_+}-
\frac{q_2^2}{p_\perp^2}\Big).\]
Furthermore
\[ A=3\qa-\frac{q_1^2}{\ra},\ \ B=4\qa,\ \ C=4\qa-2\frac{q_1^2}{\ra},\]
\[E
=-2\qa(2\ra+\rb)+r_2^2+r_1^2-k_1^2+q_1^2+2q_1^2\frac{\rb}{\ra}-
(a_1,\bar{t}-r_2)_\perp-\frac{q_1^2r_2^2}{\ra\qa}.
\]

Again we find some terms with poles at $\ra=0$ and can present
\[X_2=\tilde{X_2}+\frac{Y_2}{\ra},\ \
Y_2=-q_1^2\Big(\bar{a}+2\bar{c}_1+\bar{e}\frac{r_2^2}{\qa}+2\bar{e}p_-\Big)\]
with $\bar{c}_1=\bar{c}(\rb=-p_-)$.

We get the result
\[\Gamma_{2}=-C_2\Big(X_2^{(1)}\frac{p_+}{-\qa p_\perp^2+p_+\bar{t}_\perp^2}I_{A_2}^{(1)}
+Y_2I_{A_2}^{(2)}\frac{1}{2\qa p_-+\bar{t}_\perp^2}\Big),
\]
where $X_2^{(1)}=X_2(\ra=k_{1\perp}^2/2\qa)$ and the integrals are
\[
I_{A_2}^{(1)}=\int\frac{dr_{1-}}{2\pi}\frac{1}{k_1^2+i0}=
\int\frac{dr_{1-}}{2\pi}
\frac{1}{-2\qa r_{1-}+k_{1\perp}^2+i0}=
\frac{i}{2}\theta(-\qa)\frac{1}{\qa}
\]
and
\[I_{A_2}^{(2)}=-\frac{i}{2}\frac{1}{k_{1\perp}^2}.\]

\subsection{Fig. \ref{fig2},C}
This diagram generates two terms with different color factors.
The corresponding amplitudes $\Gamma_{3}$
and $\Gamma_{4\ep}$ are given by
\beq
\Gamma_{3,4\ep}=C_{3,4}\frac{1}{k_1^2}X_{3,4},
\eeq
where $C_3=-C_2$ and $C_4=C_1$
and the denominator is
$k_1^2=-2\qa r_{1-}+k_{1\perp}^2+i0$.

We have
\[
X_3=-(a_1\ep)_\perp
+2\frac{(p\ep)_\perp}{p_+}\Big(\frac{q_1^2}{r_{1-}}-q_{1+}\Big)-
2\frac{(p\ep)_\perp q_2^2}{p_\perp^2\rb}\Big(\frac{r_1^2}{q_{1+}}
-r_{1-}\Big)\]
and
\[
X_4=-(a_1\ep)_\perp
-\frac{(p\ep)_\perp}{p_+}\Big(\frac{q_1^2}{r_{1-}}-q_{1+}\Big)
+\frac{(p\ep)_\perp q_2^2}{p_+\ra\rb}\Big(\frac{r_1^2}{q_{1+}}
-r_{1-}\Big).\]
Here $a_i=q_i+r_i$ $i=1,2$.

As before we separate terms with poles at $\ra=0$ and $\rb=0$.
\beq
X_{3,4}=\tilde{X}_{3,4}+\frac{Y^{(1)}_{3,4}}{\ra}+\frac{Y^{(2)}_{3,4}}{\rb},
\eeq
where
\[Y^{(1)}_3=2(p\ep)_\perp\frac{q_1^2}{p_+},\ \
 Y^{(2)}_3=-2(p\ep)_\perp\frac{q_2^2}{\tp}\Big(\frac{r_1^2}{\qa}-\frac{\tp}{2p_+}\Big),\]
\[Y^{(1)}_4=(p\ep)_\perp\Big(-\frac{q_1^2}{p_+}+2\frac{q_2^2r_1^2}{\tp\qa}\Big),\ \
Y^{(2)}_4=(p\ep)_\perp\Big(-\frac{q_2^2}{p_+}+2\frac{q_2^2r_1^2}{\tp\qa}\Big).\]

The longitudinal integrals are the same for both parts and the same as for $\Gamma_2$.
From the pole at $\ra=k_{1\perp}^2/2\qa$ we get
\beq
\Gamma_{3,4}^{(1)}=C_{3,4}X_{3,4}^{(1)}I_{A_2}^{(1)},
\eeq
where $X_{3,4}^{(1)}=X_{3,4}(\ra=k_{1\perp}^2/2\qa)$.
From the poles at $\ra=0$ at $\rb=0$ we get the second contribution
\beq
\Gamma_{3,4}^{(2)}=C_{3,4}Y^{(1)}_{3,4}I_{A_2}^{(2)}-C_{3,4}Y^{(2)}_{3,4}I_{A_2}^{(2)}.
\eeq

\subsection{Fig. \ref{fig2},D}
On mass shell and convoluted with the polarization vector the corresponding amplitude $\Gamma_{5}$
is given by
\beq
\Gamma_{5}=C_5\frac{1}{k_1^2k_2^2}X_5,\ \
X_5=2(k_2L_1)L_{2}-
2(k_1L_2)L_{1}+(L_1L_2)(k_1-k_2)_\ep.
\label{a5}
\eeq
Here $k_{1,2}=q_{1,2}-r_{1,2}$,  $C_5=C_1+C_2$,
The denominators are
\[
k_2^2k_1^2=(-2\qb \rb+\tkb+i0)(-2\qa r_{1-}+k_{1\perp}^2+i0).
\]
The Lipatov vertices convoluted with polarization vectors are
\[L_1=(a_1 e)_\perp-\frac{(pe)_\perp}{p_+}\Big(\frac{q_1^2}{\ra}-\qa\Big),\ \
L_2=(a_2 e)_\perp-\frac{(pe)_\perp}{p_+}\Big(\frac{q_2^2}{\rb}-\qb\Big).\]
Also we have
\[(k_1-k_2)_\ep=(k_1-k_2,\ep)_\perp-\frac{(p\ep)_\perp}{p_+}(\qa-\qb).\]

One  finds
\beq (k_2L_1)=(pL_1)=-p_+\ra-p_-\qa+(pa_1)_\perp+r_1^2\frac{p_+}{\qa}+
q_1^2\frac{p_-}{\ra},
\eeq
\beq (k_1L_2)=(pL_2)=-p_+\rb-p_-\qb+(pa_2)_\perp+r_2^2\frac{p_+}{\qb}+
q_2^2\frac{p_-}{\rb}
\eeq
and finally
\[(L_1L_2)=(a_1a_2)_\perp+\qa\rb+\qb\ra-r_1^2\frac{\qb}{\qa}
-r_2^2\frac{\qa}{\qb}-q_1^2\frac{\rb}{\ra}-q_2^2\frac{\ra}{\rb}
+\frac{q_1^2r_2^2}{\ra\qb}+\frac{q_2^2r_1^2}{\rb\qa}.
\]

Separating the poles at $\ra=0$ and $\rb=0$ we present as before
\[X_5=\tilde{X}_5+\frac{Y_5{(1)}}{\ra}+\frac{Y_5{(2)}}{\rb}\]

Here $Y_5^{(1)}$ are sums of three terms
\[Y_5^{(1)}=
2p_-q_1^2\Big[(a_2 e)_\perp+(pe)_\perp\frac{\qb}{p_+}\Big]
\]\[
+2q_1^2\frac{(pe)_\perp}{p_+}
\Big[p_+p_--p_-\qb+(pa_2)_\perp+r_2^2\frac{p_+}{\qb}\Big]\]
\beq
+q_1^2\Big[(k_1-k_2+p,e)_\perp-2(pe)_\perp\frac{\qa}{p_+}\Big]
\Big[p_-+\frac{r_2^2}{\qb}\Big]
\eeq
and
\[Y_5^{(2)}=
-2\frac{(pe)_\perp}{p_+}q_2^2
\Big[p_+p_--p_-\qa+(pa_1)_\perp+r_1^2\frac{p_+}{\qa}\Big]
\]\[
-2p_-q_2^2\Big[(a_1e)_\perp+(pe)_\perp\frac{\qa}{p_+} \Big]\]\beq
+q_2^2\Big[(k_1-k_2-p,e)_\perp+2(pe)_\perp\frac{\qb}{p_+}\Big]
\Big[p_-+\frac{r_1^2}{\qa}\Big].
\eeq

A new longitudinal integral appears:
\beq
I_5=
\int\frac{dr_{1-}}{2\pi}\frac{1}{(2\qb
(r_{1-}-\lambda_-)+k_{2\perp}^2+i0)
(-2\qa r_{1-}+k_{1\perp}^2+i0)},
\label{i5}
\eeq
where $k_1=q_1-r_1$, $k_2=q_2-r_2$,
and we also have  contributions from residues at $\ra=0$ and
$\rb=0$.

In $I_5$ according to our rules we have to take residues in the lower half-plane.
This opens two possibilities. If both $\qa$ and $\qb$ are positive,
then only the pole at $\rb=k_{2\perp}^2/2\qb$ lies in the
lower half-plane.
If $\qb>0$ and $\qa<0$ then also the second pole at $\ra=k_{1\perp}^2/2\qa$
lies in the lower half plane.
So we find two contributions $I_5=I_5^A+I_5^B$, where
\beq
I_5^A=\frac{-i}{2}\theta(\qb)\frac{1}{\qa k_{2\perp}^2+\qb k_{1\perp}^2+2p_-\qa\qb}
\eeq
with $\rb=k_{2\perp}^2/2\qb$
and
\beq
I_5^B=-\theta(-\qa)I_5^{(1)}
\eeq
with $\ra=k_{1\perp}^2/2\qa$.

Taking into account contributions from poles at $\ra=0$ and $\rb=0$ we find
the total contribution from Fig. \ref{fig2},D
\[
\Gamma_{5}=C_5\Big\{X_5^AI_5^A
+X_5^BI_5^B-\frac{i}{2}
\Big(\frac{Y_5^{(1)}}{2\qb p_-+k_{2\perp}^2}
-\frac{Y_5^{(2}}{2\qa p_-+k_{1\perp}^2}\Big)\Big\},
\]
where $X_5^A=X_5(\rb=k_{2\perp}^2/2\qb)$ and $X_5^B=X_5(\ra=k_{1\perp}^2/2\qa)$.

\section{Appendix 2. Integration over $\kappa_+$. Light-cone poles}
To find  the final expression for  the diffractive
cross-section we have to study the eventual integration over
$\kappa_+=\qa$
that is over $\qa$ or $\qb$ with weight $\exp (iu)$, $u=zm\kappa_+/k_+$. This
integration forms function $F(z)$ according to
(\ref{ffun}). We separate it  in two parts: of the
terms which follow from the poles at $\ra=0$ or $\rb=0$ studied in this
section and of those
which follow from the Feynman poles to be studied in the next section.

The characteristic of contributions from light-cone poles is that they do not
restrict in any way the region of integration in $\kappa_+$,
which can vary from $-\infty$ to $+\infty$.

We shall study contributions with $\qa=\kappa_+$, so we
can use the formulas of the previous subsection and integrate
over $\qa$. We call the parts of the amplitudes coming from
light-cone poles as $\Gamma_i^{(2)}$, with $i=1,...5$.

In our amplitudes listed in the preceding section there appear
the following integrals over $\qa$, which follow from the residues
at $\ra=0$ or $\rb=0$.
In $\Gamma_1$ we have
\[ I_1=\int_{-\infty}^{\infty}\frac{d\qa e^{iu}}{\qa}.\]
In $\Gamma_2$ we find three integrals
\[ I_2^{(n)}=\int_{-\infty}^{\infty}\frac{\qa^n d\qa e^{iu}}{2\qa p_-+\bar{t}_\perp^2},\ \ n=1,0,-1.\]
In $\Gamma_3$ we have
\[I_3=\int_{-\infty}^{\infty}d\qa e^{iu}.\]
In $\Gamma_4$ there appear two integrals $I_3$ and $I_1$.
Finally in  $\Gamma_5$ we find two integrals
\[I_5^{(n)}=\int_{-\infty}^{\infty}\frac{\qa^n d\qa e^{iu}}{2\qa p_-+p_\perp^2-\tkb},\ \ n=1,0.\]
We only need the real parts of these integrals, since they appear
with purely imaginary coefficients after taking the residue.

In fact they are reduced to just two integrals $I_1$ and $I_3$.
Integrals $I_2^{(n)}$ with $n=1,0$ can be rewritten as
\[I_2^{(1,0)}=\frac{1}{2p_-}\int_{-\infty}^{\infty} \frac{\qa^{1,0} d\qa e^{iu}}{\qa+a_2}
=\frac{1}{2p_-}e^{-iw_2}\int_{-\infty}^{\infty} \frac{(\qa-a_2)^{1,0} d\qa e^{iu}}{\qa},\]
where
\[a_2=\frac{\bar{t}_\perp^2}{2p_-}, w_2=\frac{2mz\bar{t}_\perp^2}{M^2}<<1.\]
If we neglect $w_2$ in the exponent the two integrals reduce to $I_3$ and $I_1$.

Integrals $I_5^{(1,0)}$ are obtained from $I_2^{(1,0)}$ after substitution
$\bar{t}_\perp^2\to \tp-\tkb$.

Finally
integral $I_2^{(-1)}$ can be presented as a difference
\[I_2^{(-1)}
=\frac{1}{2p_-a_2}I_1-\frac{1}{a_2}I_2^{(0)}.\]

Now the two basic integrals are
\[I_3=2\pi\delta\Big(\frac{mz}{k_+}\Big)\propto \delta(z),\]
which does not contribute to the Glauber approximation,
and
\[I_1=i\int_{-\infty}^\infty \frac{d\qa}{\qa}
\sin\Big(\qa\frac{mz}{k_+}\Big)=i\pi\,{\rm sign}(z),\]
which is pure imaginary and odd in $z$. So it will give zero in the
amplitude.

So light-cone poles do not contribute to the diffractive amplitude.

\section{Appendix 3. Integration over $\kappa_+$: Feynman poles}
\subsection{Integration regions and problems}
Integration over $\qa$ should be done taking into account
limitations coming from the requirement that the rapidity  $y_1$ of the intermediate gluon with momentum $k_1$ or $k_2$
cannot be much different from  $y$.
Otherwise the diagram with RR$\to$RRP vertex transforms into the one with R$\to$RP vertex
with an extra interaction between the  reggeons with momenta $q_1$ and $q_2$. So
\beq
y-\delta<y_1<y+\delta,
\eeq
where one may choose $\delta>>1$ but much smaller than $\ln (s/s_0)$.
Our integrals go over negative values of $\qa$ or positive values of $\qb$.
Putting in the first case $|\qa|=p_+x$ we get the condition
\beq
x_1<x< x_2,\ \ x_{1(2)}=e^{-(+)\delta}x_0,\ \ x_0=\frac{|k_{1\perp}|}{|p_\perp|}.
\label{cond1}
\eeq
In the second case we put $\qb=p_+x$ and the integration limits in $x$ are the same
with $|k_{1\perp}|\to |k_{2\perp}|$.

Actually the integrand in the integration over $x$ contains an exponential factor
\beq
e^{-iu}, \ \  u=x\xi,\ \ \xi=-2mz\frac{\tp}{M^2}.
\label{defu}
\eeq
Note that our study is only valid when
$M^2\to \infty$, so that $\xi$  is small.

At fixed $\delta$ our integrals are naturally dependent on $\delta$. One expects that this dependence is
canceled by other contributions of the same order, which involve extra BFKL interactions between reggeons
and correction to the BFKL interaction.
Such contributions, also cut at rapidity $\delta$ from the fixed rapidity of the intermediate real gluon,
hopefully contain terms proportional to the cut integration region in rapidity, that is to $\delta$.
Thus in our calculations terms proportional to $\delta$ actually should be dropped, since they are to be
canceled by similar terms coming from other contributions. This can only be true if the divergence of our
integrals at both $x\to 0$ and $x\to\infty$ is logarithmic.

At first sight this poses the first problem in our study. Inspecting our expressions for $\Gamma_i$, $i=1,...5$ one discovers
that after integration over $\ra$ $\Gamma_{2,3,4}$ contain terms which do
not vanish at $x\to\infty$. The sum of them tend to the limit at $x\to\infty$
\[\Gamma\simeq \frac{i}{2}(pe)_\perp\frac{1}{p_+}\Big(1-2\frac{q_1^2}{k_1^2}\Big)
(-C_2-2C_3+C_4),\]
where color factors correspond to the contribution of diagrams 2,3 and 4. In the sum
$-C_2-2C_3+C_4=1/2$, so that the  limiting value is different from zero.
This generates linear divergence at large $x$ in the limit $M^2\to\infty$.
The  corresponding integral is
\beq
J_1=\int_{x_1}^{x_2} dxe^{-iu}\simeq \int_{x_1}^{x_2} dx =x_2-x_1 \ .
\label{j1}
\eeq
We shall see that after symmetrizing over participating reggeons this linear divergence is canceled.
So after integration over $\qa$ and dropping terms proportional to $\delta$ one finds the cross-section
(\ref{crsecg}).

However, this is not the end of the story.
In the result of integration over $\qa$ one finds various terms which are singular at $\tka=0$ or $\tkb=0$ and so
exhibit infrared divergence. Also we find a collinear divergence in the contribution from $\Gamma_5$.
 As we shall demonstrate, the latter cancels because of the properties of $X_5$. Terms singular at say $\tka=0$ coming
from $\Gamma_{3,4,5}$ behave  as badly as $1/|k_{1\perp}|^3$ or $\ln|k_{1\perp}|/|k_{1\perp}|^2$. However, in the sum of all
contributions
all such singular terms cancel and the remaining expression is free from the infrared divergence.

There are also terms which behave as $1/\tp$ as $\tp\to 0$, which may also lead to  divergence.
However, in our case the value of $\tp$ is limited from below by the condition that the Regge
kinematics should be valid for the lower pomerons. Fixing their minimal energy square $s_1$ as $s_1>s_0$
we find the condition
$ |\tp|>M^2s_0/2s$.
So if the terms which behave like $1/\tp$ remain in the total contribution they do not lead to infrared divergence
but rather to the
behavior $\ln (s/M^2)$ instead.
No other dangerous singularities are found in terms generated by the vertex RR$\to$RRP.

In our calculations we choose $\delta>>1$ and drop all terms proportional to $\delta$.

\subsection{ $\Gamma_1$}
We recall that after taking the residue at $\ra=\tka/(2\qa)$
\[\Gamma_1=i\frac{1}{2}C_1\theta(-\qa)\frac{1}{-\qa t_\perp^2+p_+\tka}X_1^{(1)}.\]
We find
\[X_1^{(1)}= T_0+\frac{p_+}{\qa}T_{-1},\]
where $T_{0,-1}$ do not depend on $\qa$ and
explicitly
\[T_0=-2(te)_\perp\Big(r_1^2+q_1^2+q_2^2+(a_1,t+q_2)_\perp\Big)
+2\tka(te)_\perp+4(te)_\perp\frac{q_2^2r_1^2}{\tka}\]
and
\[T_{-1}=-4(q_2e)_\perp r_1^2+2\tka (t+2r_1,e)_\perp.\]
Note that $T_0$ contains a term with a  pole at $\tka=0$. $T_{-1}$ does not contain this pole.
Integration goes over the negative $\qa$. Putting $\qa=-p_+x$ we find
\beq
\Gamma_1=-i\frac{1}{4\pi}C_1 \Big(I_0^{(1)}T_0-I_{-1}^{(1)}T_{-1}\Big),
\label{gam1a}
\eeq
where
\[I_0^{(1)}=
\int_{x_1}^{x_2}\frac{dxe^{-iu}}{x t_\perp^2+\tka},\ \
I_{-1}^{(1)}
=\int_{x_1}^{x_2}\frac{dxe^{-iu}}{x(x t_\perp^2+\tka)}\]
and  $u$ is given by (\ref{defu}).

At $x_1=0$ the integral $I_0^{(1)}$ and has a weak logarithmic  singularity at
$\tka=0$. So it is integrable and does not present difficulties itself. But $T_0$
contains a pole term in $\tka$. So the first term in (\ref{gam1a}) is in fact singular.
The second term does not exist at $x_1=0$ and is also singular at $\tka=0$.
We present
\[I_{-1}^{(1)}=\frac{1}{\tka}J_0-\frac{t_\perp^2}{\tka}I_{0}^{(1)},\]
where $J_0$ is given by
\beq
J_0=\int_{x_1}^{x_2}\frac{dx}{x} \cos(u)=\int_{\xi x_1}^{\xi x_2}\frac{du}{u}\cos u
=\ci(\xi x_2)-\ci(\xi x_1),
\label{j020}
\eeq
Here we have left only the real part, which is of interest.
As a result
\beq
\Gamma_1=-i\frac{1}{4\pi}C_1 \Big(I_{0}^{(1)}\hat{T}_0-J_0\hat{T}_{-1}\Big),
\label{gam121}
\eeq
where the new functions are
\[\hat{T}_0=
-2(te)_\perp\Big(r_1^2+q_1^2+q_2^2-\tka-t_\perp^2+(a_1,t+q_2)_\perp\Big)+4t_\perp^2(r_1e)_\perp\]
\beq
+4\frac{r_1^2}{\tka}\Big(q_2^2(te)_\perp-t_\perp^2(q_2e)_\perp\Big)
\label{ht10}
\eeq
and
\beq
\hat{T}_{-1}=\frac{1}{\tka}T_{-1}=2(t+2r_1,e)_\perp-4(q_2e)_\perp\frac{r_1^2}{\tka}.
\label{ht1}
\eeq

Note that at $k_{1\perp}=0$ we have $t_\perp=q_{2\perp}$. Therefore at $\tka=0$ the pole in $\hat{T}_0$ is
weakened to $1/|k_\perp|$. This means that the first term in (\ref{gam121}) is integrable. Singularities are separated
in the second term which contains both the singularities at $x_1=0$ and combined singularities at $x_1=0$ and
$\tka=0$. Both of them are proportional to $\delta$ and have to be discarded.

\subsection{ $\Gamma_2$}
After taking the residue at $\ra=\tka/(2\qa)$
\[\Gamma_2=-i\frac{1}{2}C_2\theta(-\qa)\frac{1}{-\qa\tp+p_+{\bar t}_\perp^2}\frac{p_+}{\qa}X_2^{(1)}.\]
One finds
\[
X_2^{(1)}=\frac{\qa^2}{p_+^2}T_2+\frac{\qa}{p_+}T_1+T_0,
\]
where $T_i$ do not depend on $\qa$ and are given by
\[T_2=-(pe)_\perp\tp\Big(1-2\frac{q_1^2}{\tka}\Big),\]
\[T_1=\bar{t}_\perp^2(pe)_\perp\Big(3-2\frac{q_1^2}{\tka}\Big)\]\[
-4\Big(p_\perp^2(r_1+r_2-p,e)_\perp+q_2^2(pe)_\perp-2(pe)_\perp (p,r_1+r_2)_\perp\Big)\]\[
+4\frac{q_1^2}{\tka}\Big(p_\perp^2(r_2-q_2,e)_\perp+(q_2^2+\tka-2(pr_2)_\perp)(pe)_\perp\Big)\]\[
+2E_0(pe)_\perp+2(q_2^2(pe)-p_\perp^2(q_2e)_\perp)\Big(1-2\frac{q_1^2}{\tka}\Big),\]
\[T_0=2\Big((q_2e)_\perp-(pe)_\perp\frac{q_2^2}{p_\perp^2}\Big)(E_0+2q_1^2).\]
Here
\beq
E_0=E(\qa=0)=r_1^2+r_2^2-q_1^2-\tka-(q_1+r_1,\bar{t}-r_2)_\perp-2\frac{q_1^2r_2^2}{\tka}.
\label{e0}
\eeq
In $T_1$ the four lines correspond to contributions from $\bar{a}A$, $(\bar{b}+\bar{c})B$,
$\bar{c}(C-B)$ and $\bar{e}E$ respectively.
All of them contain poles in $\tka=0$.

Integration over $\qa$ gives
\[\Gamma_2=-i\frac{1}{4\pi}C_2
\Big(-I_1^{(2)}T_2+I_0^{(2)}T_1-I_{-1}^{(2)}T_0\Big),\]
where
\[
I_n^{(2)}
=\int_{x_1}^{x_2}\frac{dx x^ne^{-iu}}{x\tp+\bar{t}_\perp^2}.
\]
In contrast to $\Gamma_1$ these integrals are have a finite limit at $\tka=0$, although
$I^{(2)}_{-1}$ is singular at $x_1=0$.
To separate the singularities
we present
\[
I_1^ {(2)}=\frac{1}{\tp}J_1-\frac{\bar{t}_\perp^2}{\tp}I_0^{(2)},
\ \ I_{-1}^{(2)}=\frac{1}{\bar{t}_\perp^2}J_0-\frac{\tp}{\bar{t}_\perp^2}I_0^{(2)},
\]
where $J_1$ is given by (\ref{j1})
to finally obtain
\beq
\Gamma_2=-i\frac{1}{4\pi}C_2\Big(-\frac{1}{\tp}J_1T_2-\frac{1}{\bar{t}_\perp^2}J_0T_0
+I_0^{(2)}\hat{T}_1\Big),
\label{gam21}
\eeq
with a new function
\beq
\hat{T}_1=T_1+\frac{\bar{t}_\perp^2}{\tp}T_2+\frac{\tp}{\bar{t}_\perp^2}T_0.
\label{ht21}
\eeq

Remarkably in $\hat{T_1}$, which is a sum of functions each having a pole at $\tka=0$,
the leading singularity at $\tka=0$ cancels.
Indeed the terms proportional to $q_1^2/\tka$ in (\ref{ht21}) are (modulo $q_1^2/\tka$):
from $T_2$
\[2\tp(pe)_\perp,\]
from $T_1$
\[-2\bar{t}_\perp^2(pe)_\perp+4\Big(\tp(r_2-q_2,e)_\perp+(q_2^2-2(pr_2)_\perp)(pe)_\perp\Big)
-4(q_2^2(pe)_\perp-\tp(q_2e)_\perp)-4r_2^2(pe)_\perp,\]
from $T_0$
\[-4r_2^2\Big((q_2e)_\perp-(pe)_\perp\frac{q_2^2}{\tp}\Big).\]
At $\tka=0$ we have $\bar{t}_{\perp}=-r_{2\perp}$ and $r_{2\perp}-q_{2\perp}=-p_\perp$.
Term $\tp T_2/ \bar{t}_\perp^2$ cancels the first term in $T_1$.
The second term in $T_1$ becomes
\[4(pe)_\perp\Big(q_2^2-\tp-2(pr_2)_\perp\Big)=4(pe)_\perp r_2^2\]
and cancels the last term in $T_1$.
Finally with $\bar{t}_\perp^2=r_2^2$ the term $\tp T_0/\bar{t}_\perp^2$ becomes
\[4\tp\Big((qe)_\perp-(pe)_\perp\frac{q_2^2}{\tp}\Big)\]
and cancels the remaining third term in $T_1$. So the factor multiplying
$q_1^2/\tka$ in $(\ref{ht21})$ vanishes at $\tka=0$.
Numerical studies show that in fact at
$k_{1\perp}\to 0$ this term behaves as $|k_{1\perp}|$. This means that the last term in (\ref{gam21}) is integrable
and so regular. The contributions singular at $x_1\to 0$,
$\tka\to 0$ or simultaneously in these two limits are contained in the first two terms.
As we shall see the term with $J_1$ will be eventually canceled. The second term is proportional to $\delta$
and should be dropped according to our approach. So the meaningful part of $\Gamma_2$ is contained in the third term
in (\ref{gam21}).

\subsection{$\Gamma_3$}
After taking the residue at $\ra=\tka/(2\qa)$ we have
\[\Gamma_3=iC_3\frac{1}{2\qa}\theta(-\qa)X_3,\]
Note that at $\ra=\tka/(2\qa)$ we have
\[\rb=\frac{\qa\tp-p_+\tka}{2p_+\qa},\ \ \frac{1}{\rb}=\frac{2\qa p_+}{d},\ \ d=\qa\tp-p_+\tka.
\]
We find
\beq
X_3=\frac{\qa}{p_+}T_1+T_0+\frac{p_+}{d}V_0,
\eeq
where
\[T_1=-2(pe)_\perp\Big(1-2\frac{q_1^2}{k_{1\perp}^2}\Big),\]
\[T_0=-(a_1e)_\perp,\]
\beq
V_0=2(pe)_\perp(k_{1\perp}^2-2r_1^2)\frac{q_2^2}{p_\perp^2}.
\label{v03}
\eeq
Integration over  $\qa=-xp_+$ gives
\beq
\Gamma_3=i\frac{1}{4\pi}C_3\Big(T_1J_1-T_0J_0+I_{-1}^{(3)}V_0\Big).
\label{gam3}
\eeq
Here $J_1$ and $J_0$ are the old integrals which do not depend on any variables and
$I_{-1}^{(3)}$ is new:
\[I_{-1}^{(3)}=
\int_{x_1}^{x_2}\frac{dxe^{-iu}}{x(x\tp+\tka)}=
\frac{1}{\tka}J_0-\frac{\tp}{\tka}I_0^{(3)},\]
where
\[
I_0^{(3)}=\int_{x_1}^{x_2}\frac{dxe^{-iu}}{x\tp+\tka}.
\]
So we finally have
\beq
\Gamma_3=i\frac{1}{4\pi}C_3\Big[T_1J_1+J_0\Big(-T_0+\frac{1}{\tka}V_0\Big)-I_0^{(3)}\frac{\tp}{\tka}V_0\Big].
\label{gam31}
\eeq

The integral $I_0^{(3)}$ converges at $x_1=0$ and has a weak (logarithmic) singularity
at $\tka=0$. It is integrable over $k_{1\perp}$ by itself but $V_0$ contains a pole term
in $\tka$. So dropping the term with $J_1$, apart from the trivial singular term with $J_0$
we get a strongly singular term
\beq
\Gamma_3^{sing}=iC_2\frac{1}{\pi}(pe)_\perp\frac{q_2^2r_1^2}{\tka}I^{(3)}_0,
\label{gan3s}
\eeq
which is not integrable in $k_{1\perp}$.

\subsection{$\Gamma_4$}
After taking the residue at $\ra=\tka/(2\qa)$ we have
\[\Gamma_4=iC_4\frac{1}{2\qa}\theta(-\qa)X_4^{(1)},\]
where $X_4^{(1)}=X_4(\ra=\tka/(2\qa))$.

So we find
\beq
X_4^{(1)}=\frac{\qa}{p_+}T_1+T_0+\frac{\qa}{d}V_1,
\eeq
where
\[T_1=(pe)_\perp\Big(1-2\frac{q_1^2}{\tka}\Big),\]
\[T_0=-(a_1e)_\perp,\]
\beq
V_1=-2(pe)_\perp(\tka-2r_1^2)\frac{q_2^2}{\tka},
\label{v14}
\eeq

Integration over $\qa=-xp_+$ gives
\beq
\Gamma_4=i\frac{1}{4\pi}C_4\Big(J_1T_1-J_0T_0-I^{(3)}_0V_1\Big).
\label{gam41}
\eeq
Dropping the term with $J_1$ we find a trivial term with $J_0$ and a highly singular term
similar to that in $\Gamma_3$
\beq
\Gamma_4^{sing}=-\frac{i}{\pi}C_4(pe)_\perp\frac{q_2^2r_1^2}{\tka}I_0^{(3)}.
\label{gam4s}
\eeq
It differs from a similar term in $\Gamma_3$ only by the sign. However, since
$C_3/C_4=2$ these terms do not cancel in the sum.

\subsection{$\Gamma_5^{(A,B)}$}

{\bf 1. Case A}: $\rb=\tkb/2\qb$ ($\qb>0$).

We express $\qa=p_+-\qb$. We then have
\[\ra=\frac{\tp}{2p_+}-\frac{\tkb}{2\qb},\ \ \frac{1}{\ra}=\frac{2p_+\qb}{\qb\tp-p_+\tkb}\equiv
\frac{2p_+\qb}{d_2}.\]
The common denominator is
\[D=p_+\tkb+\qb (\tka-\tkb-\tp)-2p_-\qb^2.\]
Putting $\qb=xp_+$ we rewrite it as $D=p_+R_2$ with $ R_2=(k_{2\perp}-xp_\perp)^2$.

As before we present
\beq
X_5^{(A)}=x^2T_2+xT_1+T_0+T_{-1}\frac{1}{x}+U_0\frac{1}{\qa}+V_0\frac{1}{d_2},
\label{xaix}
\eeq
where
\[ T_2=\tp\alpha_2(pe)_\perp,\]
\[T_1=-\tp(a_2e)_\perp-\tp\alpha_2(a_1+p,e)_\perp+\frac{1}{2}\tp\alpha_2(\kappa_2e)_\perp
+2(pe)_\perp\Big(\alpha_2(pa_2)_\perp+(pa_2)_\perp-c_2\Big),\]
\[T_0=2(a_2e)_\perp\Big((pa_1)_\perp-q_1^2\Big)-\alpha_2b_{12}(pe)_\perp-2(a_1+p,e)_\perp(pa_2)_\perp\]\[
+4(pe)_\perp(pa_2)_\perp\frac{q_1^2}{\tp}+c_2(\kappa_2e)_\perp+2\alpha_2r_1^2(pe)_\perp
+2b_{22}(pe)_\perp\frac{q_1^2}{\tp},\]
\[T_{-1}=\tkb(a_2e)_\perp-2b_{22}(q_1e)_\perp,\]
\[U_0=p_+r_1^2\Big(2(a_2e)_\perp-\alpha_2(\kappa_2e)_\perp\Big),\]
\beq
V_0=-2p_+\tkb q_1^2(a_2e)_\perp+p_+b_{22}q_1^2(\kappa_2e)_\perp
+2p_+b_{22}q_1^2(pe)_\perp\Big(1+\frac{\tkb}{\tp}\Big)+4p_+q_1^2(pa_2)_\perp(pe)_\perp\frac{\tkb}{\tp}.
\label{tuva}
\eeq
In these formulas
\[\alpha_2=1-2\frac{q_2^2}{\tkb},\ \ b_{ij}=2r_i^2-k_{j\perp}^2,\ \ \kappa_2=k_1-k_2-p\]
and
\[c_2=(a_1a_2)_\perp+r_1^2+r_2^2+q_2^2-\tkb-2\frac{q_2^2r_1^2}{\tkb}.\]

{\bf 2. Case B}: $\ra=\tka/2\qa$ ($\qa<0$).

We express $\qb=p_+-\qa$. We then have
\[\rb=\frac{\tp}{2p_+}-\frac{\tka}{2\qa},\ \ \frac{1}{\rb}=\frac{2p_+\qa}{\qa\tp-p_+\tka}\equiv
\frac{2p_+\qa}{d_1}.\]
The denominator is
\[D=p_+\tka+\qa (\tkb-\tka-\tp)-2p_-\qa^2\]
Putting $\qa=xp_+$ with $x<0$ we rewrite the denominator as $D=p_+R_1$ with
$R_1=(k_{1\perp}-xp_\perp)^2$

Presenting again
\beq
X_5^{(B)}=x^2\tilde{T}_2+x\tilde{T}_1+\tilde{T}_0+\tilde{T}_{-1}\frac{1}{x}+\tilde{U}_0\frac{1}{\qb}+\tilde{V}_0\frac{1}{d_1}
\label{xbix}
\eeq
we find
\[ \tilde{T}_2=-\tp\alpha_1(pe)_\perp.\]
\[
\tilde{T}_1=-2(pe)_\perp\Big((pa_1)_\perp+\alpha_1(pa_2)_\perp+(a_1a_2)_\perp-\tka+r_1^2+q_1^2
-\alpha_1r_2^2\Big)
\]\[+\alpha_1\tp(a_2+p,e)_\perp+\tp(a_1e)_\perp+\frac{1}{2}\alpha_1(\kappa_1e)_\perp,
\]
\[
\tilde{T}_0=(pe)_\perp\Big(\alpha_1b_{21}+2b_{11}\frac{q_2^2}{\tp}-4(pa_1)\frac{q_2^2}{\tp}\Big)\]
\[
+2(a_2+p,e)_\perp (pa_1)_\perp+2(a_1e)_\perp(pa_2)_\perp+2q_1^2(a_1e)_\perp
+(\kappa_1e)_\perp (a_1a_2)_\perp,
\]
\[
\tilde{T}_{-1}=2b_{11}(q_2,e)_\perp-\tka(a_1e)_\perp,
\]
\[
\tilde{U}_0=-p_+r_2^2\Big(\alpha_1(\kappa_1e)_\perp+2(a_1e)_\perp\Big),
\]
\beq
\tilde{V}_0=2p_+q_2^2\tka(a_1e)_\perp+p_+b_{11}q_2^2\Big((\kappa_1e)_\perp-2(pe)_\perp\Big)
-2p_+q_2^2(pe)_\perp\frac{\tka}{\tp}\Big(b_{11}+2(pa_1)_\perp\Big).
\label{tuvb}
\eeq
In these formulas
\[\alpha_1=1-2\frac{q_1^2}{\tka},\ \ b_{ij}=2r_i^2-k_{j\perp}^2,\ \ \kappa_1=k_1-k_2+p.\]

We pass to the final integration of $\Gamma_5$ over $\qa$ or $\qb$.

{\bf Integration of $X_5^A$.}

We have to integrate over the transferred "+"-momentum $\kappa_+$, which is
equivalent to integration over $\qa$ or $\qb$. In case A it is convenient to integrate
over $\qb=xp_+$ with $x>0$. Before integration
\[\Gamma_5^A=C_5X_5^{(A)}F_5^A,\]
where
\[ F_5^A=-\frac{i}{2}\theta(\qb)\frac{1}{\qa\tkb+\qb\tka+2p_-\qa\qb}.\]
The denominator can be written in Euclidean metric as
\[\qa\tkb+\qb\tka+2p_-\qa\qb=-p_+D_2,\ \ D_2=({\bf k}_2-x{\bf p})^2.\]
Presenting $X_5^{(A)}$ according to Eq. (\ref{xaix}) we shall have the integrated $\Gamma_5^A$ as
\beq
\Gamma_5^A=\frac{i}{4\pi}C_5\Big(\sum_{n=-1}^2T_{n}I^A_n+\frac{U_0}{p_+}K_1^A-\frac{V_0}{p_+}K^A_2\Big),
\label{gama}
\eeq
where
\[I^A_n=\int_{x_1}^{x_2}\frac{x^n dx}{(k_2-xp)^2},\]
\[K_1^A=\int_{x_1}^{x_2}\frac{dx}{(1-x)(k_2-xp)^2},\]
\[K_2^A=\int_{x_1}^{x_2}\frac{ dx}{(xp^2-k_2^2)(k_2-xp)^2},\]
and all vectors are to be taken in Euclidean 2-dimensional metric (E2DM) with $k_2^2,p^2\geq 0$.
All integrals are well convergent at small $x$ so that we can safely put $x_1=0$.
At large $x$ $I^A_n$ with $n=1,2$ diverge and we have to separate the diverging terms.
All integrals are standard and apart from these divergent terms they are expressed via
integral
\beq
I^A_0=\int_{x_1}^{x_2}\frac{ dx}{(k_2-xp)^2}=\frac{\pi-\phi_2}{\sqrt{p^2k_2^2-(pk_2)^2}},
\label{ia0}
\eeq
where $\phi_2$ is the angle between ${\bf k}_2$ and ${\bf p}$, $0<\phi_2<\pi$.
All the rest integrals are given as follows
\[ I^A_1=\frac{1}{p^2}\delta+\frac{(pk_2)}{p^2}I^A_0,\]
\[I^A_2=\frac{1}{p^2}J_1+2\frac{(pk_2)}{p^2}I^A_1-\frac{k_2^2}{p^2}I^A_0,\]
\[I^A_{-1}=\frac{1}{k_2^2}\delta+\frac{(pk_2)}{k_2^2}I^A_0,\]
\[K^A_1=-\frac{1}{2k_1^2}\ln\frac{k_2^2}{p^2}+\frac{(pk_1)}{k_1^2}I^A_0,\]
\[K_2^A=-\frac{1}{2k_1^2k_2^2}\ln\frac{k_2^2}{p^2}+\frac{(k_1k_2)}{k_1^2k_2^2}I^A_0.\]
Here ${\bf k}_1={\bf p}-{\bf k}_2$. Integral $J_1$ is given by (\ref{j1}).

The basic integral $I^A_0$ contains both a collinear singularity at $\phi_2=0$ and
a comparatively weak (integrable)  singularity at $k_2\to 0$ where it behaves as $1/k_2$.
 Note that the denominator in $I^A_0$ is symmetric in $k_1$ and $k_2$;
\beq
k_1^2p^2-(pk_1)^2=k_2^2p^2-(pk_2)^2=k_1^2k_2^2-(k_1k_2)^2,
\label{denn}
\eeq
so that $I^A_0$ also contains a similar singularity at $k_1\to 0$

The collinear singularity is in fact absent, since the whole $X^A$ vanishes when $k_1$ and $k_2$\
are directed along $p$. Indeed recall that $k_{2+}=\qb=xp_+$ and $k_{2-}=-\rb$. Then $\rb=\tkb/2\qb$
implies that $k_2^2=0$, that is the gluon $k_2$ lies on its mass-shell. The collinear singularity
occurs at $k_\perp=xp_\perp$ from which we find $k_{2\perp}/k_+=p_\perp/p_+$. Taking its square we have
\[
\frac{\tkb}{k_{2+}^2}=\frac{p_\perp^2}{p_+^2}\ \ {\rm or}\ \ \frac{k_{2-}}{k_{2+}}=\frac{p_-}{p_+}.
\]
The last equation means $k_{2-}=xp_-$, so that the 4-vector $k_2=xp$. Since $p=k_1+k_2$ then
$k_1=(1-x)p$ and both $k_1$ and $k_2$ are parallel to $p$. But for such $k_1$ and $k_2$ $X_5=0$,
since (in the 4-dimensional Lorentz metric)
\[(L_1k_2)=x(L_1p)=\frac{x}{1-x}(L_1k_1)=0,\ \ (L_2,k_1)=\frac{1-x}{x}(L_2k_2)=0\]
and
\[(k_1-k_2,e)=(1-2x)(pe)=0.\]

Dropping the divergent term with $J_1$ on the reasons discussed earlier we present the integrated $\Gamma^A$ as a sum of two terms,
one of which is proportional to $\ln k_2^2$ and the other proportional to $I^A_0$:
\[
\Gamma^A=\frac{i}{4\pi}C_5\Big\{-\frac{1}{2p_+k_1^2}\Big(U_0-\frac{1}{k_2^2}V_0\Big)\ln\frac{k_2^2}{p^2}+\]\beq+
I^A_0\Big(\Big(2\frac{(pk_2)^2}{p^4}-\frac{k_2^2}{p^2}\Big)T_2+\frac{(pk_2)}{p^2}T_1+T_0+\frac{(pk_2)}{k_2^2}T_{-1}
+\frac{(pk_1)}{p_+k_1^2}U_0-\frac{(k_1k_2)}{p_+k_1^2k_2^2}V_0\Big)\Big\}.
\label{gama1}
\eeq

Our aim is to analyze possible non-integrable divergencies of this expression in the limit $k_2\to 0$.
The power-like singularities come both from  singularities in integrals $I^A_n$, $K_1^A$ and $K_2^A$
and from the coefficients $T_n$, $U_0$ and $V_0$. Obviously term with $T_2$ is non-singular due to
factor $(pk_2)^2$. Also non-singular terms in $T_1$, $T_0$ and $U_0$  do not lead to non-integrable singularity
as $k_2\to 0$.  So dangerous terms include terms of the order $1/k_2^2$ in $T_1$, $T_0$  and terms non-vanishing as $k_2\to 0$
in $T_{-1}$, $U_0$ and $V_0$.

We start from $T_1$. Terms of order $1/k_2^2$ come from coefficient $\alpha_2=1-2q_2^2/\tkb$. Collecting them we find
in the limit $k_2\to 0$ (in E2DM)
\[T_1=-4\frac{q_2^2}{k_2^2}\Big(q_{1}^2(pe)-p^2(q_1e)\Big).\]
Next we find that  terms proportional to $1/k_2^2$ are in fact absent in $T_0$ and $U_0$.
We are left with $T_{-1}$ and $V_0$. In the limit $k_2\to 0$ (also in E2DM)
\[ T_{-1}=-4r_2^2(q_1e),\ \
V_0=-4p_+r_2^2q_1^2(pe).\]
The singular terms from $T_1$, $T_{-1}$ and $V_0$ give in the sum
\[4\frac{(pk_2)q_2^2}{k_2^2}\Big(p^2(q_1e)-q_1^2(pe)\Big)+4\frac{r_2^2(pk_2)}{k_2^2}\Big(q_1^2(pe)-p^2(q_1e)\Big).\]
At $k_2=0$ we have $q_2=r_2$ so that all power-like singularities cancel.

So the only dangerous terms at $k_2\to 0$ are the logarithmic ones. They come only from the contribution proportional to $V_0$
and lead to the singular term (in Lorentz metric)
\beq
\Gamma^A_{k_2\to 0}=\frac{i}{\pi}C_5\frac{q_1^2r_2^2(pe)_\perp}{2p_\perp^2}\,\frac{1}{\tkb}\ln\frac{\tkb}{p_\perp^2}.
\label{gamas}
\eeq

To finish with case A we consider behavior at $k_1\to 0$. Obviously in this case only terms with $U_0$ and $V_0$ may lead
to non-integrable singularities. So we have to calculate values of $U_0$ and $V_0$ at $k_1\to 0$ and so $k_2\to p$.
Elementary calculations give in E2DM
\[U_0=4p_+r_1^2\Big((q_2e)-\frac{q_2^2}{p^2}(pe)\Big),\ \
V_0=p^2U_0.\]
As a result the singular terms proportional to $(pk_1)/k_1^2$ cancel, so that $\Gamma^A$ is integrable at $k_1=0$.
Its only singularity is at $k_2\to 0$ and given by (\ref{gamas})

{\bf Integration of $X_5^B$.}

Case B is considered quite similarly.
Now we integrate over $\qa=xp_+$ with $x<0$. Before integration
\[\Gamma_5^B=C_5X_5^{(B)}F_5^B,\]
where
\[ F_5^B=\frac{i}\theta(-\qa)\frac{1}{\qa\tkb+\qb\tka+2p_-\qa\qb}.\]
The denominator can be written in Euclidean metric as
\[\qa\tkb+\qb\tka+2p_-\qa\qb=-p_+D_1,\ \ D_1=({\bf k}_1-x{\bf p})^2.\]
Presenting $X_5^{(B)}$ according to Eq. (\ref{xbix}) we shall have the integrated $\Gamma_5^B$
as
\beq
\Gamma_5^B=-\frac{i}{4\pi}C_5\Big(\sum_{n=-1}^2\tilde{T}_{n}I^B_n+\frac{\tilde{U}_0}{p_+}K_1^B-\frac{\tilde{V}_0}{p_+}K_2^B\Big),
\label{gamb}
\eeq
where tildes mean case $B$. The integrals are
\[I^B_n=\int_{-x_2}^{-x_1}\frac{x^n dx}{(k_1-xp)^2},\]
\[K_1=\int_{-x_2}^{-x_1}\frac{dx}{(1-x)(k_1-xp)^2},\]
\[K_2=\int_{-x_2}^{-x_1}\frac{ dx}{(xp^2-k_1^2)(k_1-xp)^2}\]
and all vectors are  taken in E2DM.
All integrals are well convergent at small $x_i\to 0$ so that we van safely put $x_1=0$.
At large $|x|$ $I^B_n$ with $n=1,2$ diverge and we have to separate the diverging terms.
All integrals  are expressed via
integral
\beq
I^B_0=\int_{-x_2}^{-x_1}\frac{ dx}{(k_1-xp)^2}=\frac{\phi_1}{\sqrt{p^2k_1^2-(pk_1)^2}},
\label{ib0}
\eeq
where $\phi_1$ is the angle between ${\bf k}_1$ and ${\bf p}$, $0<\phi_1<\pi$.
All the rest integrals are given as follows
\[ I^B_1=-\frac{1}{p^2}\delta+\frac{(pk_1)}{p^2}I^B_0,\]
\[I^B_2=\frac{1}{p^2}J_1+2\frac{(pk_1)}{p^2}I^B_1-\frac{k_1^2}{p^2}I^B_0,\]
\[I^B_{-1}=-\frac{1}{k_1^2}\delta+\frac{(pk_1)}{k_1^2}I^B_0,\]
\[K^B_1=\frac{1}{2k_2^2}\ln\frac{k_1^2}{p^2}+\frac{(pk_2)}{k_2^2}I^B_0,\]
\[K_2^B=\frac{1}{2k_1^2k_2^2}\ln\frac{k_1^2}{p^2}+\frac{(k_1k_2)}{k_1^2k_2^2}I^B_0.\]

The denominator in $I^B_0$ is  the same as in $I^A_0$ (see (\ref{den})), the difference between them
is in the numerator.
As in case A the basic integral $I^B_0$ contains both a collinear singularity at $\phi_1=\pi$ and
comparatively weak (integrable)  singularities at $k_1\to 0$ and $k_2\to 0$ where it behaves as $1/k_1$
or $1/k_2$. As mentioned, the collinear singularity is in fact absent, since the whole $X^B$ vanishes when $k_1$ and $k_2$
are parallel to $p$.
Dropping the divergent term with $J_1$ on the reasons discussed earlier we present the integrated $\Gamma^B$ as a sum of two terms,
one of which is proportional to $\ln k_1^2$ and the other proportional to $I^B_0$:
\[
\Gamma^B=-\frac{i}{4\pi}C_5\Big\{\frac{1}{2p_+k_2^2}\Big(\tilde{U}_0-\frac{1}{k_1^2}\tilde{V}_0\Big)\ln\frac{k_1^2}{p^2}+\]\beq+
I^B_0\Big[\Big(2\frac{(pk_1)^2}{p^4}-\frac{k_1^2}{p^2}\Big)\tilde{T}_2+\frac{(pk_1)}{p^2}\tilde{T}_1+\tilde{T}_0+\frac{(pk_1)}{k_1^2}\tilde{T}_{-1}
+\frac{(pk_2)}{p_+k_2^2}\tilde{U}_0-\frac{(k_1k_2)}{p_+k_1^2k_2^2}\tilde{V}_0\Big)\Big\}.
\label{gamb2}
\eeq

We first analyze possible non-integrable singularities as $k_1\to 0$.
The power-like singularities come again from  singularities of integrals $I^B_n$, $K_1^B$ and $K_2^B$
and from the coefficients $\tilde{T}_n$, $\tilde{U}_0$ and $\tilde{K}_0$. As in case A term with $\tilde{T}_2$ is non-singular due to
factor $(pk_1)^2$. Also non-singular terms in $\tilde{T}_1$, $\tilde{T}_0$ and $\tilde{U}_0$  do not lead to non-integrable singularity
as $k_1\to 0$.  So as in case A, dangerous terms include terms of the order $1/k_1^2$ in $\tilde{T}_1$, $\tilde{T}_0$  and terms non-vanishing as $k_1\to 0$
in $\tilde{T}_{-1}$, $\tilde{U}_0$  and $\tilde{V}_0$

We start from $\tilde{T}_1$. Terms of order $1/k_1^2$ come from coefficient $\alpha_1=1-2q_1^2/\tka$. Collecting them we find
in the limit $k_1\to 0$ (in E2DM)
\[\tilde{T}_1=4\frac{q_1^2}{k_1^2}\Big(q_{2}^2(pe)-p^2(q_2e)\Big).\]
Next we find that terms proportional to $1/k_1^2$ are in fact absent in $\tilde{T}_0$ and $\tilde{U}_0$.
We are left with $\tilde{T}_{-1}$ and $\tilde{V}_0$. In the limit $k_1\to 0$ in E2DM
\[\tilde{T}_{-1}=4r_1^2(q_2e),\ \
\tilde{V}_0=4p_+r_1^2q_2^2(pe).\]
The singular terms from $\tilde{T}_1$, $\tilde{T}_{-1}$ and $\tilde{V}_0$ give in the sum
\[-4\frac{(pk_1)q_1^2}{k_1^2}\Big(p^2(q_2e)-q_2^2(pe)\Big)-4\frac{r_1^2(pk_1)}{k_1^2}\Big(q_2^2(pe)-p^2(q_2e)\Big).\]
At $k_1=0$ we have $q_1=r_1$ so that all power-like singularities cancel.

So the only dangerous terms at $k_1\to 0$ are the logarithmic ones. They come only from the contribution proportional to $\tilde{V}_0$
and lead to the singular term (in Lorentz metric)
\beq
\Gamma^B_{k_1\to 0}=-\frac{i}{\pi}C_5\frac{q_2^2r_1^2(pe)_\perp}{2p_\perp^2}\,\frac{1}{\tka}\ln\frac{\tka}{p_\perp^2}.
\label{gambs}
\eeq

Finally we consider behavior at $k_2\to 0$. In  this case only terms with $\tilde{U}_0$ and $\tilde{V}_0$ may lead
to non-integrable singularities. So we have to calculate values of $\tilde{U}_0$ and $\tilde{V}_0$ at $k_2\to 0$ and so $k_1\to p$.
Calculations give in E2DM
\[\tilde{U}_0=-4p_+r_2^2\Big((q_1e)-\frac{q_1^2}{p^2}(pe)\Big),\ \
\tilde{V}_0=p^2\tilde{U}_0 .\]
So the singular terms proportional to $(pk_2)/k_2^2$ cancel, and  $\Gamma^B$ turns out to be integrable at $k_2=0$.
Its only singularity is at $k_1\to 0$ and given by (\ref{gambs}).

\subsection{Final singularities}

We first consider the dangerous non-integrable logarithmic singularity at $k_1\to 0$ which comes
from $\Gamma_3$, $\Gamma_4$ and $\Gamma^B$.
As we have found from $\Gamma_3$ and $\Gamma_4$ in the sum the singular term is
\[
\Gamma^{sing}_{3+4}=\frac{i}{\pi}(C_3-C_4)(pe)_\perp\frac{q_{2\perp}^2r_{1\perp}^2}{\tka}\int_{x_1}^{x_2}\frac{dx}{xp_\perp^2+\tka}.
\]
Passing to E2DM we rewrite this as
\[
\Gamma^{sing}_{3+4}=-\frac{i}{\pi}(C_3-C_4)(pe)\frac{q_{2}^2r_{1}^2}{k_1^2}\int_{x_1}^{x_2}\frac{dx}{xp^2+k_1^2}.
\]
The integral gives
\[
\int_{x_1}^{x_2}\frac{dx}{xp^2+k^2}=\frac{1}{p^2}\ln\frac{x_2p^2+k_1^2}{x_1p^2+k_1^2}.
\]
At this point we recall that $x_1=(k_1/p)e^{-\delta}$ and $x_2=(k_1/p)e^\delta$ with $\delta>>1$.
Then we find
\[
\int_{x_1}^{x_2}\frac{dx}{xp^2+k^2}=\frac{1}{p^2}\ln\frac{pke^\delta}{k_1^2}=-\frac{1}{2p^2}\ln\frac{k_1^2}{p^2}
\]
and the singular term from $\Gamma_{3+4}$ becomes
\beq
\Gamma^{sing}_{3+4}=\frac{i}{\pi}(C_3-C_4)(pe)\frac{q_{2}^2r_{1}^2}{2p^2k_1^2}\Big(\ln\frac{k_1^2}{p^2}-2\delta\Big).
\label{s34}
\eeq

Now we pass to the singular contribution from $\Gamma^B$ given by (\ref{gambs}).
It was obtained by taking $x_1=0$. However, it does not change if one takes $x_1=(k_1/p)e^{-\delta}$ instead
(unlike $I^{(3)}_0$ , due to convergence at high $x_2$). The logarithm in (\ref{gambs}) comes in fact from
\[
\ln\frac{(k_1^2+x_1p^2)^2}{(\bk_1+x_1\bp)^2}-(x_1\to x_2)=
\ln\frac{(k_1^2+k_1pe^{-\delta})^2}{(\bk_1+(k_1/p)\bp e^{-\delta})^2}-(\delta\to -\delta).\]
In the limit $\delta>>1$ this goes into $\ln (k_1^2/p^2)$, which is the same if we took $x_1=0$ and $x_2\to\infty$
from the start.

In E2DM (\ref{gambs}) reads
\beq
\Gamma^{sing}_B=\frac{i}{\pi}C_5\frac{q_2^2r_1^2(pe)}{2p^2k_1^2}\ln\frac{\tka}{p_\perp^2}.
\label{sb}
\eeq
Dropping the term with $\delta$ in(\ref{s34}) and summing with (\ref{sb})we find the final logarithmic singularity
\beq
\Gamma^{sing}_{3+4+B}=\frac{i}{\pi}(C_3-C_4+C_5)(pe)\frac{q_2^2r_1^2}{2p^2k_1^2}\ln\frac{k_1^2}{p^2}.
\eeq
Recalling that $C_3=-1$, $C_4=-1/2$ and $C_5=1/2$ we find that this singularity is canceled.

Thus the only singular term in the whole integrated $\Gamma$ are those proportional either to $J_1$ or to $\delta$.
Hopefully the latter  can safely be discarded due to cancelations with similar terms from the diagrams with pairwise interregional interaction.
Terms with $J_1$ actually go after symmetrization, as will be demonstrated below.

\subsection{Terms with $J_1$}
These terms are present in linear diverging expressions in $\Gamma_i$. $i=2,...5$
The corresponding contributions from $\Gamma_{2,3,4}$ are
\[\Gamma_2=\frac{1}{4\pi}C_2J_1\frac{1}{p_\perp^2}T_2^{(2)},\ \ T_2^{(2)}=-(pe)_\perp\tp\alpha_1,\]
\[\Gamma_3=\frac{1}{4\pi}C_3J_1T_1^{(3)},\ \ T_1^{(3)}=-2(pe)_\perp\alpha_1,\]
\[\Gamma_4=\frac{1}{4\pi}C_4J_1T_1^{(3)},\ \ T_1^{(3)}=(pe)_\perp\alpha_1,\]
where $\alpha_1=1-2q_1^2/\tka$.
In the sum we get
\beq
\Gamma_{2+3+4}=\frac{1}{4\pi}\alpha_1(pe)_\perp J_1(-C_2-2C_3+C_4)=\frac{1}{8\pi}\alpha_1(pe)_\perp J_1.
\eeq
We have to add to this contribution the three others obtained by interchanges $(q_1\lra q_2)$ and $(r_1\lra r_2)$.
At these exchanges both longitudinal and transversal momentum components are to be exchanged. Exchange $q_1 \lra q_2$  does not change
the momentum part since one has to integrate both over $q_{1,2+}$ and $q_{1,2\perp}$. However, this exchange gives a minus sign due to
change of the color factor. Exchange $r_1\lra r_2$ does not change the color factor but we have to take residues in the
lower half-plane of $\ra$. This means upper half plane of variable $\rb$. Therefore simultaneously with changing
$r_{1\perp}\lra r_{2\perp}$ after taking the residue one should change $-i\theta(-\qa)\lra i\theta(\qb)$, which is equivalent
to taking $x\to -x$ in our formulas plus the overall minus sign. In practice this means that in terms with $J_1$ we have to take
the minus sign.
Denoting the result of adding the interchanged contribution as Sym we get
\beq
{\rm Sym}\,\Gamma_{2+3+4}=-\frac{i}{8\pi}J_1(pe)_\perp \Big(\frac{q_1^2}{(q_1-r_1)_\perp^2}-\frac{q_2^2}{(q_2-r_1)_\perp^2}
-\frac{q_1^2}{(q_1-r_2)_\perp}+\frac{q_2^2}{(q_2-r_2)^2}\Big).
\label{t1j1}
\eeq
This expression should be integrated with the pomeron symmetric in $r_{1\perp}$ and $r_{2\perp}$. Then the result of integration
vanishes. So in the end no terms proportional to $J_1$ come from $\Gamma_{2+3+4}$.
This means that the sum of these vertices has no linear divergency after symmetrization.

From $\Gamma_5$ we have two contributions:
\[\Gamma^A=\frac{i}{4\pi}C_5J_1\frac{1}{|\tp|}T_2,\ \ T_2=-|\tp|\alpha_2(pe)_\perp,\]
\[\Gamma^B=-\frac{i}{4\pi}C_5J_1\frac{1}{|\tp|}\tilde{T}_2,\ \ \tilde{T}_2=|\tp|\alpha_1(pe)_\perp.\]
In the sum we find
\beq
\Gamma_5=\frac{i}{2\pi}C_5J_1(pe)_\perp \Big(\frac{q_2^2}{\tkb}+\frac{q_1^2}{\tka}-1\Big).
\label{g5ab}
\eeq
Symmetrization can be done either by changing $q_1\lra q_2$ or changing $r_1\lra r_2$. In both cases we
have the sign minus. So
\beq
{\rm Sym}\,\Gamma^5=\frac{i}{2\pi}C_5J_1(pe)_\perp \Big(\frac{q_2^2}{(q_2-r_2)_\perp^2}+\frac{q_1^2}{(q_1-r_1)_\perp^2}
-\frac{q_2^2}{(q_2-r_1)_\perp^2}-\frac{q_1^2}{(q_1-r_2)_\perp^2}\Big).
\label{t2j}
\eeq
This expression is antisymmetric in $r_{1\perp}$ and $r_{2\perp}$ and so vanishes after integration with the pomeron.

Thus the whole contribution from the vertex $\Gamma$ does not contain terms proportional to $J_1$ and so does not contain  linear divergence.

\end{document}